\documentclass[aps,floatfix,preprint,showpacs,preprintnumbers,nofootinbib,superscriptaddress,natbib]{revtex4}

\usepackage[titletoc]{appendix}
\usepackage{graphicx,float}
\usepackage{epsfig}		
\usepackage{dcolumn}
\usepackage{bm}
\usepackage{subfigure}
\usepackage{verbatim}
\usepackage[all]{xy}
\usepackage{amsmath,upgreek}
\usepackage{amssymb}

\usepackage{pdfpages}
\usepackage{color}
\usepackage{graphicx,epstopdf}
\usepackage[colorlinks=true,
 linkcolor=red,
 urlcolor=purple,
 citecolor=blue,hyperindex]{hyperref}

\def\0{\mbox{\tiny $0$}}
\def\1{\mbox{\tiny $1$}}
\def\2{\mbox{\tiny $2$}}
\def\3{\mbox{\tiny $3$}}
\def\4{\mbox{\tiny $4$}}
\def\5{\mbox{\tiny $5$}}
\def\6{\mbox{\tiny $6$}}
\def\7{\mbox{\tiny $7$}}
\def\8{\mbox{\tiny $8$}}
\def\9{\mbox{\tiny $9$}}

\def\f14{\mbox{\tiny $\frac{1}{4}$}}

\begin{document}
	\title{Intrinsic quantum correlations for Gaussian localized Dirac cat states in phase space}
\author{Caio Fernando e Silva}
\email{caiofernandosilva@df.ufscar.br}
\author{Alex E. Bernardini}
\email{alexeb@ufscar.br}
\affiliation{~Departamento de F\'{\i}sica, Universidade Federal de S\~ao Carlos, PO Box 676, 13565-905, S\~ao Carlos, SP, Brasil.}
	
	\date{\today}
	
\begin{abstract}
Following the information-based approach to Dirac spinors under a constant magnetic field, the phase-space representation of symmetric and anti-symmetric localized Dirac cat states is obtained.
The intrinsic entanglement profile implied by the Dirac Hamiltonian is then investigated so as to shed a light on quantum states as carriers of qubits correlated by phase-space variables. 
Corresponding to the superposition of Gaussian states, cat states exhibit non-trivial elementary information dynamics which include the interplay between intrinsic entanglement and quantum superposition as reported by the corresponding Dirac archetypes. Despite the involved time-evolution as non-stationary states, the Wigner function constrains the elementary information quantifiers according to a robust framework which can be consistently used for quantifying the time-dependent $SU(2) \otimes SU(2)$ (spin projection and intrinsic parity) correlation profile of phase-space localized Dirac spinor states. Our results show that the Dirac Wigner functions for cat states -- described in terms of generalized Laguerre polynomials -- exhibit an almost maximized timely persistent mutual information profile which is engendered by either classical- or quantum-like spin-parity correlations, depending on the magnetic field intensity.
	\end{abstract}
	
	\date{\today}
	\maketitle
\section{Introduction}	

Quantum superposition and entanglement are two intertwined concepts that lie at the very heart of quantum mechanics.
Considering a broader description of the informational aspects of two-qubit systems \cite{Vedral,Henderson,n024,Auyuanet}, formal extensions to include continuous degrees of freedom and quantify their influence on localization-decoherence and correlation information aspects have been investigated in the last decades. It includes, for instance, a plethora of issues closer to magneto-electronics \cite{Prinz} and spintronics \cite{todos}, to the evaluation of quantum computation techniques \cite{Xi,Ma,Chitambar}, and to the reported existence of cat states in quantum optics \cite{Bermudez}.
In particular, experimental platforms for cat states \cite{cat1,cat2,cat3,cat4,cat5} driven by Dirac spinor structures have already featured subtle properties of non-classical phenomena, and the setup for continuous variable dependent Dirac spinor solutions driving the elementary information content of such systems must be investigated.
In the scope of the Dirac-like systems \cite{extfields,diraclike01,PRB001,PRB002,PRA2018,BernardiniCP}, which also includes experimental platforms for low-energy and mesoscopic phenomena \cite{n001,n002,Gerritsma,Bermudez,n005,n006,graph03,graph04,CastroNeto}, confining potentials implemented by Dirac Hamiltonians have only been recently addressed as the evolution operator for a two-qubit system codified by position and momentum variables \cite{BernardiniEPJP}. 
According to the spinor structure framework developed for describing spinor associated intrinsic quantum correlations \cite{extfields,diraclike01,PRB001,PRB002,PRA2018}, the intrinsic spin-parity correlations of Dirac spinors were classified in terms of Poincar\'e classes of Dirac constant potentials \cite{extfields}. 
At the same time, concerning the inclusion of continuous degrees of freedom, the Weyl-Wigner phase-space formalism \cite{Wigner,Moyal,Ballentine} has been cast into the spinor structure form of a Dirac-Wigner formalism  \cite{1983,1986,1987,BernardiniEPJP} so as to include position and momentum variables as the drivers of quantum correlations.

Since the $SU(2) \otimes SU(2)$ group structure of the Dirac equation encompasses the systematicness for probing intrinsic entanglement \cite{PRA2018}, a non-trivial entanglement profile from the interference between localized Dirac spinors is expected. It is supported by the $SU(2) \otimes SU(2)$ algebra related to the spin-parity degrees of freedom, which, in turn, are generally correlated by position and momentum degrees of freedom as described by their dynamics \cite{BernardiniCP}. 
Concomitantly, in the Weyl-Wigner framework \cite{Wigner,Moyal,Ballentine}, the Wigner function can be expanded around the classical probability distribution in phase space \cite{Huang}, thus becoming a quantumness quantifier \cite{Kenfack} and allowing for the understanding of quantum-to-classical transitions. Furthermore, the Weyl-Wigner approach evinces a quantum informational perspective on localized quantum states, as the Wigner function suggests a straightforward connection to probability distributions. Considering that Dirac spinors also exhibit an information-based structure associated to Hilbert spaces of finite dimensions, it is relevant to comprehend the interplay between resource-based spinors and the phase-space representation. As a matter of fact, given the above-mentioned group structure which drives the Dirac equation alongside the spinor decomposition, the Wigner function itself can be related to the density operator of an information theory for confined particles, from which quantum correlations can be quantified algebraically \cite{1983,1986,1987,BernardiniEPJP}.

The Wigner function decomposition into the sixteen generators of the corresponding Clifford algebra \cite{1986} clarifies the connection between the Wigner representation and the standard formulation of quantum mechanics, even according to the second quantization framework. Indeed, an explicit one-to-one correspondence between these two approaches can be established so as to provide, for instance, the theoretical tools for the computation of the quantum purity. In particular, it can be obtained either from the aforementioned decomposition or from the coordinate representation of spin-parity traced out density matrix, from which the analysis of more generic Wigner functions for a dynamical system can be performed.

Exploring the implications of superpositions on the information profile of confined Dirac spinors, the dynamics of a fermion under a magnetic field shall be investigated. In particular, the formation of Landau levels described by Laguerre polynomials in phase space shall be recovered so as to allow one to derive the non-trivial dynamics of two relevant configurations of quantum states. Firstly, Dirac-Gaussian states shall be engendered from a superposition involving only Dirac eigenstates with the same principal quantum number. 
Besides their easy-working mathematical properties, and due to their correspondence with their classical counterparts, Gaussian states usually work as an effective measurement platform, for instance, in quantum optics experiments \cite{Sch} and in the scope of quantum chemistry \cite{Stewart,Faegri,Huzinaga,Hill,Hill2} involving molecular integration techniques \cite{Guseinov1979,Guseinov1985,Weniger}. Considering that Gaussian Wigner functions are non-negative definite, and conversely, that the reported existence of Gaussian engendered cat states may result from the interference between different Landau levels driven by the above mentioned Dirac Hamiltonian dynamics \cite{BernardiniEPJP}, Dirac cat states shall be explicitly described both in configuration and phase spaces. Through this second frame, Dirac cat states are suited for interpreting and clarifying the properties of non-classical phenomena, with their corresponding elementary information content being analytically computed. 

With the final aim of obtaining the intrinsic entanglement profile implied by the Hamiltonian for a charged fermion trapped by a magnetic field ${\mathbf B}$, described according to the Dirac spinor structure, Gaussian and cat states described as superpositions of associated Dirac spinor stationary states are engendered and read as carriers of qubits correlated by phase-space variables. Such Dirac Wigner functions for cat states -- once described in terms of generalized Laguerre polynomials -- provide the elements for the evaluation of spin-parity correlations depending on the magnetic field intensity, which is the final goal of this manuscript.

The paper is thus organized as follows. In section II, from initial Gaussian superpositions, symmetrical and anti-symmetrical Dirac cat states are obtained for fermions described as Dirac spinors under a magnetic field. In section III, the phase space Wigner formalism for Dirac spinors is briefly recovered in order establish the grounds for quantifying local and global spin-parity correlations for the localized states introduced in section II. Analytical tools for obtaining phase space averaged information quantifiers, namely for quantum purity and mutual information, are implemented.
More relevantly, considering the phase-space dynamical evolution and the Gaussian pattern of the involved systems, the measure of the Dirac spin-parity non-separability is obtained in terms of the associated quantum concurrence, which is computed in a two-fold way: $i)$ as the difference between total and classical mutual information between continuous and discrete degrees of freedom implied by the Dirac equation; and $ii)$ from the previous formulation applied to two-qubit quantum systems, now applied to localized states. Our conclusions are drawn in section IV, where the main findings of our work are summarized and relevant extensions are posed to future investigation.

\section{Dynamics of Dirac localized states in configuration space} \label{sec2}

The stationary states for a charged fermion trapped by a magnetic field ${\bf B} $ can be obtained from the dynamical evolution driven by the Hamiltonian,
\begin{equation}
H = \mbox{\boldmath$\alpha$} \cdot ({\bf p} + (-1)^r\, e {\bf A}) + \beta m,
\label{eq0}
\end{equation}
where the potential vector, ${\bf A}$, results into the magnetic field ${\bf B} = \mbox{\boldmath$\nabla$} \times {\bf A}$, $e$ is the positive unit of charge, and $r = 1$ and $2$ label the positive and negative intrinsic parity states, respectively. For the gauge chosen as ${\bf A} = \mathcal{B}\,x\, \hat{\bf y}$, which corresponds to a magnetic field along the $z$-direction, a set of orthogonal Dirac Hamiltonian eigenstates from \eqref{eq0} can be written as \cite{BernardiniEPJP}
\begin{equation}
\psi = \exp\big[i(-1)^r E_n t + k_y y + k_z z\big] u_{n,r} ^\pm (s_r), \label{stationarysolutions}
\end{equation}
i.e. plane-wave solutions in both $y$ and $z$ directions. For compactness of the notation, the parameters $A_n$, $B_n$, and $\eta_n$,
\begin{equation} \label{parameters}
A_n = \frac{k_z}{E_n + m}, \quad B_n = \frac{\sqrt{2n\, e \mathcal{B}}}{E_n + m}, \quad \eta_n = \frac{E_n + m}{2E_n},
\end{equation}
are introduced for describing the energy associated parameters so as to resume a set of constraints given by $0 \leq A_n$, $B_n \leq 1$ and $\eta_n(A_n ^2 + B_n ^2 + 1) =1$, for the energy of the $n$-th Landau level identified by
\begin{equation}
(-1)^r E_n = (-1)^r \sqrt{m^2 + k_z ^2 + 2n e \mathcal{B}},\quad \mbox{with}\quad r = 1,\,2.
\end{equation}
To summarize the influence of the magnetic field, the dynamics along the $x$-coordinate is shifted according to
\begin{equation}
s_r = \sqrt{e {\mathcal B}} \left( x + (-1)^r {k_y \over e{\mathcal B}}
\right),
\label{222}
\end{equation}
such that the positive parity ($r=1$) space-dependent spinors can be written as
\begin{eqnarray}\label{9998}
u^+_{n,1}(s_1) = \sqrt{\eta_{n}}\left( \begin{array}{c} 
\mathcal{F}_{n-1}(s_1) \\ 0 \\ 
A_{n}\, \mathcal{F}_{n-1}(s_1) \\
-B_{n}\, \mathcal{F}_{n} (s_1) 
\end{array} \right), \quad 
u^-_{n,1}(s_1) = \sqrt{\eta_{n}}\left( \begin{array}{c} 
0 \\ \mathcal{F}_{n} (s_1) \\
-B_{n}\,
\mathcal{F}_{n-1}(s_1) \\ 
-A_{n}\,\mathcal{F}_{n}(s_1)
\end{array} \right), \quad 
\end{eqnarray}
as well as the negative parity ($r=2$) ones as
\begin{eqnarray}
u^+_{n,2}(s_2) = \sqrt{\eta_{n}}\left( \begin{array}{c} 
B_{n}\,
\mathcal{F}_{n-1}(s_2) \\ 
A_{n}\,\mathcal{F}_{n}(s_2) \\ 
0 \\ \mathcal{F}_{n} (s_2)
\end{array} \right), \qquad 
u^-_{n,2}(s_2)= \sqrt{\eta_{n}}\left( \begin{array}{c} 
-A_{n}\,
\mathcal{F}_{n-1}(s_2) \\ 
B_{n}\,
\mathcal{F}_{n} (s_2) \\ 
\mathcal{F}_{n-1}(s_2) \\ 0
\end{array} \right),
\label{9999}
\end{eqnarray}
where the functions $\mathcal{F}_n (s_r)$ are related to the Hermite polynomials, $H_n (s_r)$, by
\begin{equation}
\mathcal{F}_n (s_r) = \left( \frac{\sqrt{e \mathcal{B}}}{n! \, 2^n \sqrt{\pi}}\right)^{1/2} e^{-(s_r) ^2/2} H_n (s_r),
\end{equation}
which are only defined for non-negative integers $n$ and imply into the following properties,\footnote{A possible definition for negative integers $l$ is simply $\mathcal{F}_{l} (s_r)= 0$.}
\begin{equation}
\int ds \, \mathcal{F}_n (s) \mathcal{F}_m (s) = \sqrt{e \mathcal{B}} \, \delta_{mn},
\end{equation}
and
\begin{equation}
\sum_n \mathcal{F}_n (s) \mathcal{F}_n (s') = \sqrt{e \mathcal{B}}\, \delta (s - s') = \delta (x - x'),
\end{equation}
i.e. the orthonormalization and completeness relations, respectively. An equivalent basis of eigenfunctions was used in \cite{Canuto,Proskurin}. 

At this point, it is worth to mention that the definition of the $s_r$-coordinate in Eq.~\eqref{222} takes into account the intrinsic parity and momentum orientation of the plane wave solutions, allowing one to easily implement the orthogonality relations between spinors.\footnote{The compact expression for the $s_r$-coordinate and the spin polarization of spinors $u ^\pm _{n,2}$ should be clear when one works with the negative parity states; thus, the physical spin operator is also defined with opposite sign. In the language of the hole theory, this corresponds to redefining the spin projection for fermions with negative energy \cite{Greiner}. Previously, one has worked with the stationary solutions only, so reversing momentum is harmless; here, one is interested in non-stationary states, and thus the momentum sign must be carefully chosen. 
} For instance,
\begin{equation}
\int ds \, {u_{n,1} ^\pm} (s_1) ^\dagger {u_{n,1} ^\mp} (s_1) = \int ds \, {u_{n,2} ^\pm} (s_2) ^\dagger {u_{n,2} ^\mp} (s_2) = 0.
\end{equation}
Therefore, spinors with the same parity but opposite spin projection are orthogonal. These relations can be extended to spinors with opposite parity by noticing that
\begin{equation}
(u^+ _{n,1}(s_1))^\dagger u^+ _{n,2}(s_2) = \eta_n B_n \bigg( \mathcal{F}_{n-1}(s_1) \mathcal{F}_{n-1}(s_2) - \mathcal{F}_{n}(s_1) \mathcal{F}_{n}(s_2) \bigg),
\end{equation} 
which does not vanish upon integration due to the distinct arguments inside the functions. One can either reverse the momentum of the negative parity states or simply set $k_y = 0$ so that $s_1 = s_2 = s$, since changing the momentum of the corresponding plane wave is not desirable. In this way, orthogonality relations become
\begin{equation}
\int ds \, {u_{n,1} ^\pm (s) } ^\dagger {u_{n,2} ^\pm (s) } = \int ds \, {u_{n,1} ^\pm (s) } ^\dagger {u_{n,2} ^\mp (s)} = 0, \label{spinornormalization}
\end{equation}
from which non-stationary states can finally be engendered.

Suppressing the arguments by setting $\mathcal{F}_n (s) \equiv \mathcal{F}_n$, $u_{n,1} ^\pm (s) \equiv u_{n,1} ^\pm$, and so forth, the following superposition of eigenstates is proposed, 
\begin{eqnarray} \label{Gaussianstate1}
\mathcal{G}_{n} ^{^{(1)}} (s,\,t) &=& \bigg(\exp(-i E_n t)u_{n,1} ^+ + \exp(i E_n t)(- A_n u_{n,2} ^- + B_n u_{n,2} ^+) \bigg) \nonumber \\
&=& \sqrt{\eta_n} \left\{ \exp(-i E_n t)\left( \begin{array}{cc}
\mathcal{F}_{n-1}\\ 
0 \\
A_n \mathcal{F}_{n-1} \\
- B_n \mathcal{F}_{n}\\ \end{array} \right) + 
 \exp(i E_n t) \left( \begin{array}{cc}
(B_n ^2 + A_n ^2 )\mathcal{F}_{n-1} \\ 
0 \\
-A_n \mathcal{F}_{n-1} \\
+B_n \mathcal{F}_{n} \\ \end{array} \right) \nonumber \right \} \\
&\equiv& \eta_n
\left( \begin{array}{cc} \label{spinor1}
\bigg[\exp(-i E_n t) + \exp(i E_n t) (A_n ^2 + B_n ^2)] \bigg] \mathcal{F}_{n-1} \\ 
0 \\
 -2 i \sin(E_n t) A_n \mathcal{F}_{n-1} \\
 2 i \sin(E_n t) B_n \mathcal{F}_{n}\\ 
\end{array} \right).
\end{eqnarray}\normalsize
One notices that the $y$ and $z$ exponential dependent term was omitted, since all waves travel with the same momentum, and thus the relevant $1$-dim dynamics is along the $s$-coordinate.

The states above described by $\mathcal{G}_{n} ^{^{(1)}} (s,\,t)$ exhibit a simple form for $t=0$, $\left(\begin{smallmatrix} \mathcal{F}_{_{n-1}} & 0 & 0& 0 \end{smallmatrix}\right)^T$ such that, if $n=1$, one has a Gaussian state,
\begin{equation}\label{eq9}
\mathcal{F}_0 (s) = \left( \frac{\sqrt{e \mathcal{B}}}{2 \sqrt{\pi}}\right)^{1/2} e^{-s^2/2},
\end{equation}
which simply corresponds to the lowest Hermite polynomial.

A complete basis, in the sense of spinor components, can be obtained with distinct polarizations, for instance, as 
\begin{eqnarray}
\mathcal{G}_{n} ^{^{(2)}} (s,\,t) &=& \bigg(\exp(-iE_n t) u_{n,1} ^- + \exp(iE_n t)( A_n u_{n,2} ^+ + B_n u_{n,2} ^-) \bigg) \nonumber \\
& \equiv & \eta_n
\left( \begin{array}{cc} \label{spinor2}
0 \\ 
\bigg[\exp(-i E_n t) + \exp(i E_n t) (A_n ^2 + B_n ^2)] \bigg] \mathcal{F}_{n}\\
 2 i \sin(E_n t) B_n \mathcal{F}_{n-1} \\
 2 i \sin(E_n t) A_n \mathcal{F}_{n}\\ 
\end{array} \right),
\end{eqnarray}
and the two remaining spinors are similarly obtained as
\begin{eqnarray}
\mathcal{G}_n ^{^{(3)}} (s,\,t) &=& \bigg(\exp(iE_n t) u_{n,2} ^- + \exp(-iE_n t) ( A_n u_{n,1} ^+ - B_n u_{n,1} ^- ) \bigg)
\end{eqnarray}
and
\begin{eqnarray}
\mathcal{G}_n ^{^{(4)}} (s,\,t) &=& \bigg( \exp(-iE_n t)(- B_n u_{n,1} ^+ - A_n u_{n,1} ^+) + \exp(iE_n t) u_{n,2} ^+ \bigg), \label{444}\end{eqnarray}
which encompass the four time-dependent quantum states that can describe departing Gaussian states with distinct spin-parity polarizations. However, setting $n=0$ in $\mathcal{G}_n ^{^{(2)}} $ and $\mathcal{G}_n ^{^{(4)}} $ yields states without relevant dynamics, since the spatial part will permanently be Gaussian for any $t$. They can be contrasted with $\mathcal{G}_n ^{^{(1)}} $ and $\mathcal{G}_n ^{^{(3)}} $, which can be prepared as an initial Gaussian state for $n=1$ that evolves into a non-Gaussian state due to the contribution from $\mathcal{F}_1 (s)$. As it shall be depicted in the following, when the phase-space formulation is considered, the choice of the particular polarization has implications onto the local aspects of the quantum information content.

\subsection{Cat states}

Before moving on to the Wigner formalism, the quantum states obtained above can also be worked out so as to encompass the interference between states with non-coincident quantum numbers. Generically, from the generalized quantum superposition given by
\begin{equation}\label{superposition}
\phi _{i} (s,\,t) = \mathcal{N}^{1/2} \sum^{\infty}_{n=0} c_n \, \mathcal{G}_{n} ^{(i)} (s,\,t).
\end{equation}
with the normalization constant $\mathcal{N}$, and with $\mathcal{G}_{n} ^{(i)}$ obtained from Eqs.~\eqref{Gaussianstate1}-\eqref{444}
for $i =1,\,2,\,3,\,4$, one has, for instance, for $i=1$, $c_{2n+1}=0$ and $c_{2n} = \exp(-a^2/4) (a/\sqrt{2}) ^{2n} / \sqrt{(2n)!}$, with $a$ parameterizing a dimensionless distance, the only non-vanishing component of the Dirac spinor for $t=0$ given by $ \left(\begin{matrix} 1 & 0 & 0& 0 \end{matrix}\right)^T$ multiplied by
\begin{equation}\label{simp}
 \exp(-a^2/4) \,\sum_{n=0} ^{\infty} \mathcal{F}_ {2n} (s) \frac{(a/\sqrt{2})^{2n}}{\sqrt{(2n)!}} = \left( \frac{e \mathcal{B}}{\pi}\right)^{1/4} e^{-s^2/2} \sum_{n=0} ^{\infty} \frac{ H_{2n} (s) }{(2n)!} \left(\frac{a}{2}\right)^{2n}.
\end{equation}
Since one has the even contributions from the infinite sum from Eq.~\eqref{superposition}, the expression from \eqref{simp} simplifies into \cite{Gradshteyn}
\begin{equation}\label{catstate}
\phi^S (s, t=0) = \frac{1}{2}\left( \frac{e \mathcal{B}}{\pi} \right)^{1/4} \bigg \{ \exp\left[-\frac{1}{2}(s-a)^2 \right] + \exp\left[-\frac{1}{2}(s+a)^2 \right] \bigg \} \left(\begin{matrix} 1 & 0 & 0& 0 \end{matrix}\right)^T,
\end{equation}
where the index $S$ stands for a symmetric superposition of two Gaussian states centered at $s = \pm a$: a symmetric Dirac cat state. Including the time time-dependent factors from Eq.~\eqref{spinor1}, the time-evolved $S$-state is written as 
\small
\begin{equation}\label{SCS}
\phi^S (s,\,t) = \begin{pmatrix} 
\phi^S_1 (s,\,t) \\
0 \\
\phi^S_3 (s,\,t)\\
\phi^S_4(s,\,t)
\end{pmatrix},
\end{equation}
\normalsize
with
\begin{eqnarray}
\phi^S _1 (s,\,t) &=& \sum _{n=0} ^{\infty} \frac{e^{-a^2/4}\mathcal{F}_{2n} (s)}{1+ A^{2} _{2n+1} + B^{2} _{2n+1} } \frac{(a/\sqrt{2})^{2n}}{\sqrt{(2n)!}} \bigg( e^{-i E_{2n+1}t} + (A^{2} _{2n+1} + B^{2} _{2n+1}) e^{i E_{2n+1}t} \bigg) \label{qstte1}, \\
\phi^S _3 (s,\,t) &=& - 2i \sum _{n=0} ^{\infty} \frac{e^{-a^2/4}A_{2n+1}}{1+ A^{2} _{2n+1} + B^{2} _{2n+1} } \frac{(a/\sqrt{2})^{2n}}{\sqrt{(2n)!}} \mathcal{F}_{2n} (s) \sin( E_{2n+1}t) \label{qstte2}, \\
\phi^S _4 (s,\,t) &=& 2i \sum _{n=0} ^{\infty} \frac{e^{-a^2/4} B_{2n+1}}{1+ A^{2} _{2n+1} + B^{2} _{2n+1} } \frac{(a/\sqrt{2})^{2n}}{\sqrt{(2n)!}} \mathcal{F}_{2n+1} (s) \sin( E_{2n+1}t).\label{qstte3} 
\end{eqnarray}

Analogously, anti-symmetric (A) cat states can be engendered from the odd contributions from the infinite sum from Eq.~\eqref{superposition}, i.e. by setting $c_{2n}=0$ and $c_{2n+1} = \exp(-a^2/4) (a/\sqrt{2}) ^{2n+1} / \sqrt{(2n+1)!}$. Following the same procedure, the initial spinor becomes 
\begin{equation}
\phi^A (s,t=0) = \frac{1}{2} \left( \frac{e \mathcal{B}}{\pi} \right)^{1/4} \bigg \{ \exp\left[-\frac{1}{2}(s-a)^2 \right] - \exp\left[-\frac{1}{2}(s+a)^2 \right] \bigg \} \left(\begin{matrix} 1 & 0 & 0& 0 \end{matrix}\right)^T,
\end{equation}
and the time-evolved $A$-state can thus be written in the general form of 
\small
\begin{equation}\label{ACS}
\phi^A (s,\,t) = \begin{pmatrix} 
\phi^A_1 (s,\,t) \\ 
0 \\
\phi^A_3 (s,\,t)\\
\phi^A_4(s,\,t)
\end{pmatrix},
\end{equation}
\normalsize
with
\begin{eqnarray}
\phi^A _1 (s,\,t) &=& \sum _{n=1} ^{\infty} \frac{e^{-a^2/4}\mathcal{F}_{2n-1} (s)}{1+ A^{2} _{2n} + B^{2} _{2n} } \frac{(a/\sqrt{2})^{2n-1}}{\sqrt{(2n-1)!}} \bigg( e^{-i E_{2n} t} + (A^{2} _{2n} + B^{2} _{2n}) e^{i E_{2n}t} \bigg), \\
\phi^A _3 (s,\,t) &=& - 2i \sum _{n=1} ^{\infty} \frac{e^{-a^2/4}A_{2n}}{1+ A^{2} _{2n} + B^{2} _{2n} } \frac{(a/\sqrt{2})^{2n-1}}{\sqrt{(2n-1)!}} \mathcal{F}_{2n-1} (s) \sin( E_{2n}t), \\
\phi^A _4 (s,\,t) &=& 2i \sum _{n=1} ^{\infty} \frac{e^{-a^2/4} B_{2n}}{1+ A^{2} _{2n} + B^{2} _{2n} } \frac{(a/\sqrt{2})^{2n-1}}{\sqrt{(2n-1)!}} \mathcal{F}_{2n} (s) \sin( E_{2n}t).
\end{eqnarray}
Of course, similar cat states could be initialized with different polarizations, by replacing $\mathcal{G}_n ^{(1)}$ by $\mathcal{G}_n ^{(2,3,4)}$ into Eq.~\eqref{superposition}. 

Just to sum up, although the above quantum states were obtained in terms of an infinite sum composition, normalization and purity conditions shall impose additional constraints that simplify the algebraic manipulations involving them. Besides, contrarily to the previous Gaussian wave functions, which are usually regarded as the closest classical realizations of particles, cat states have an explicit entanglement profile \cite{Karimipuor,Shen}.  
Hence, their intrinsic information profile, and how it is affected by the quantum superposition evolution shall be evaluated with the support of the Wigner phase-space framework.

\section{Time-dependent information profile of coherent superpositions in phase space} \label{sec3}

The description of localization under confining potentials is akin to the Wigner approach for both non-relativistic and relativistic quantum mechanics. The covariant matrix-valued Wigner function mapped by the covariant Dirac equation structure \cite{1983,1986,1987,BernardiniEPJP} indeed supports a decomposition in terms of the sixteen generators of the Clifford algebra, $\{\gamma_{\mu},\gamma_{\nu}\} = 2g_{\mu\nu}$. However, the covariance is lost due to the presence of the magnetic field which, however, is accommodated by the definition of the equal-time Wigner function for a fixed reference frame \cite{BernardiniEPJP}. Recalling the Dirac representation, for which the gamma matrices are given by $\gamma_{0} = \beta$, $\gamma_{j} = \beta\alpha_j$, $\{\gamma_{\mu},\gamma_{5}\} = 0$, and $\sigma_{\mu\nu} = (i/2)[\gamma_{\mu},\gamma_{\nu}]$, the Wigner function can be decomposed as \cite{Weickgenannt}
\begin{equation}\label{51}
\omega(\{q\}) \equiv
\mathcal{S}(\{q\})+
i\,\gamma_{5}\,{\Pi}(\{q\})+
\gamma_{\mu}\,\mathcal{V}^{\mu}(\{q\})+
\gamma_{\mu}\gamma_{5}\,\mathcal{A}^{\mu}(\{q\})+
\frac{1}{2}\sigma_{\mu\nu}\mathcal{T}^{\mu\nu}(\{q\}),
\end{equation}
with $\{q\}\equiv \{\mathbf{x},\,\mathbf{k};\,t\}$. Multiplying the left-hand side by the corresponding generator that appears in front of each term and tracing over spinorial indices, the scalar, pseudo-scalar, vector, axial-vector, and anti-symmetric tensor contributions are all correspondently identified \cite{1983,1986,1987}. 

Moving to the computation of the Wigner function from a particular spinor configuration, the Weyl transform can be applied to the relevant density operator. Thus, the phase-space dynamics can be described by the equal-time Dirac-like Wigner function \cite{Zhuang,Sheng} that supports the aforementioned decomposition.\footnote{From now on, the Wigner function employed refers to the equal-time expression, instead of the covariant one.} A superposition of stationary states can be generally put into the following form 
\begin{equation}
\phi_\lambda (x + u ) = \sum _{j} \psi_{\lambda,j} (\bm{x} + \bm{u})\exp[-i k_{0,j} (t + \tau)],
\end{equation}
for $t$, $\tau$ and $k_0$ the time-like components of $x$, $u$ and $k$, where the index $j$ simply labels the $j$-th spinor in the superposition for a particular orthonormalized basis. Then, the Wigner function can be computed as
\begin{eqnarray}
\omega_{ \xi \lambda} (\bm{x},\bm{k};t)&=& \int ^{+\infty} _{-\infty} \hspace{-1em} d \mathcal{E} \, W_{\lambda \xi} (x,k) \nonumber \\
&=& \pi^{-1} \sum _{j,m} \exp[ i(k_{0,j} - k_{0,m} ) t] \int d\tau \int ^{+\infty} _{-\infty} \hspace{-1em} d \mathcal{E} \exp[ -i( 2\mathcal{E} - k_{0,j} - k_{0,m} ) \tau] \nonumber \\
&& \quad \times \quad \pi ^{-3} \int d^3 \bm{u} \exp[2i \bm{k}. \bm{u}] \bar{\psi}_{\lambda,j}(\bm{x} - \bm{u})\psi_{\xi,m}(\bm{x} + \bm{u}) \nonumber \\
&=& \pi^{-3} \sum _{j,m} \exp[ i(k_{0,j} - k_{0,m}) t] \int d^3 \bm{u} \exp[2i \bm{k}. \bm{u}] \bar{\psi}_{\lambda,j}(\bm{x} - \bm{u})\psi_{\xi,m}(\bm{x} + \bm{u}),\label{stationarywignerfunction}
\end{eqnarray}
where the last row is obtained by evaluating the integrals over $\tau$ and then $\mathcal{E}$. The above definition is understood as an energy-averaged Wigner function for a fixed frame; furthermore, it does not equal the sum of the Wigner functions corresponding to stationary states, given that the linearity of the Dirac equation is lost when moving to the phase space. 

The normalization of the probability distribution is obtained by setting $\lambda = \xi$ and integrating over phase space, i.e.
\begin{equation}
\int d^{3}\mathbf{x}\, \int d^{3}\mathbf{k} \, Tr\left[\gamma_{0}\,\omega_{\xi\lambda}(\mathbf{x},\,\mathbf{k};\,t)\right] = \mathcal{N},
\end{equation}
where the trace operation is over spinorial indices, and $\mathcal{N}$ only depends on the coefficients of the superposition if the quantum states are orthonormalized; if there is a single (stationary) state, it follows that $\mathcal{N}=1$. This is a generalization of the Schr\"{o}dinger-like Wigner function \cite{Wigner} that incorporates the $SU(2)\otimes SU(2)$ group structure associated to Dirac spinors into the Weyl-Wigner phase-space formalism. From the same perspective, an extension to statistical mixtures is also possible, with the quantum purity for Dirac spinors simply generalized to \cite{BernardiniEPJP}
\small\begin{eqnarray}
\mathcal{P}&=& 8\pi^3\int\hspace{-.2cm}d^3\mathbf{x}\int\hspace{-.2cm}d^3\mathbf{k} \, Tr\left[\left(\gamma^{0}\omega(\mathbf{x},\,\mathbf{k};\,t)\right)^2\right] = 
8\pi^3\int\hspace{-.2cm}d^3\mathbf{x}\int\hspace{-.2cm}d^3\mathbf{k} \, Tr\left[\omega(\mathbf{x},\,\mathbf{k};\,t)\,\omega^{\dagger}(\mathbf{x},\,\mathbf{k};\,t)\right],
\end{eqnarray}\normalsize
where the extra factor of $8 \pi^3$ ensures the pure-state constraint as $\mathcal{P}=1$. Of course, this is a straightforward extension of the purity expression for non-relativistic quantum mechanics, $Tr[\hat{\rho}^2]$, once the density matrix is identified with the Wigner function via the Weyl transform of quantum operators \cite{Case,BernardiniEPJP}. In both cases, the quantum purity quantifies the loss of information that can be usually associated to system-environment interactions such as thermalization effects on quantum fluctuations \cite{Silva}.

Information quantifiers associated to continuous and discrete degrees of freedom are calculated by means of the purity expression applied to the corresponding Hilbert space. To clear up this assertion, the relative linear entropies related to spin-parity and phase-space coordinates are
\begin{eqnarray}\label{linlin}
\mathcal{I}_{SP} &=& 1 - Tr\left[\left(\langle\omega_{\xi\lambda}\rangle \gamma_0\right)^2\right]
,\end{eqnarray}
and
\begin{eqnarray}
\mathcal{I}_{\{\mathbf{x},\,\mathbf{k}\}} &=& 1 - (2\pi)^3 
\int\hspace{-.2cm}d^3\mathbf{x}\int\hspace{-.2cm}d^3\mathbf{k}\,\left(Tr\left[\omega_{\xi\lambda} (\mathbf{x},\,\mathbf{k};\,t) \gamma_0\right]\right)^2,
\end{eqnarray}
respectively. The brackets in the first expression indicate phase-space averaging, in a correspondence to a trace operation over continuous degrees of freedom. Conversely, $Tr[...]$ is always understood as a trace over discrete indices, which averages out the spin-parity subspace. 

As in standard information theory, the mutual information between spin-parity and phase-space degrees of freedom can be calculated from the entropies above and amounts to the total correlation between discrete and continuous degrees of freedom,
\begin{equation}\label{mut}
M^{SP} _{x, k_x} = \mathcal{I}_{\{x,k_x\}} + \mathcal{I}_{SP} + \mathcal{P}- 1,
\end{equation}
which can be both classical- and quantum-like. If the corresponding Hilbert spaces coexist independently, mutual information vanishes and, eventually, classical and quantum mutual correlations are distinguished.

For more engendered configurations involving, for instance, electron correlation effects in molecular structures
where quantum mutual information between orbitals are evaluated \cite{Rissler,Tecmer,Giribe,Ding},
mutual information follows from the strict seminal connection with von Neumann (vN) entropies, $\mathcal{S}^{(vN)}$, which replace the linear entropies at Eq.~(\ref{mut}), so as to return
\begin{equation}\label{mut2}
M^{SP(vN)}_{x, k_x} = \mathcal{S}^{(vN)}_{\{x,k_x\}} + \mathcal{S}^{(vN)}_{SP} - \mathcal{S}^{(vN)}_{Tot},
\end{equation}
In fact, for peaked phase space distributions as Gaussian states, it can be demonstrated that $\mathcal{S}^{(vN)}\approx 1-\mathcal{P}$ with highly sufficient confidence level \footnote{The simplest approach for computing the quantum entropy content of the Wigner function can be achieved introducing an additional contribution to $\mathcal{S}^{(vN)}$ given by $-\ln(2\pi)$, that is:
\begin{eqnarray}
\mathcal{S}^{(vN)} = -\ln(2\pi) - \int_{V}dV\, W\,\ln(W) &=& -\int_{V}dV\, W\,\ln(2\pi \,W) 
= \int_{V}dV\, W - 2\pi \int_{V}dV\, W^2 + \dots\nonumber\\ &=& 1 - \mathcal{P} + (\mbox{higher order terms}).
\end{eqnarray}.}. In this case, according to Eq.~\eqref{linlin}, a straightforward connection of the Wigner {\em quasi}-probability distribution correspondence with the density matrix interpretation is enabled.

The above introduced tools will be applied to give a broader understanding of the information carried by the Dirac Gaussian and cat states previously obtained. In particular, time-dependent mutual information and quantum entanglement will be analytically given in terms of the external field, $\mathcal{B}$. 

\subsection{Fermionic Gaussian state dynamics under a magnetic field}

For the Gaussian state, $\mathcal{G}_n ^{^{(1)}} $ (cf. Eq.~\eqref{Gaussianstate1}), the corresponding Wigner function can be computed from the matrix multiplication $\mathcal{G}_n ^{^{(1)}}(\mathcal{G}_n ^{^{(1)}})^\dagger \gamma_0$, 
\small
\begin{equation} \label{wignermatrix}
\omega _{n,1} (s,k_x;t) = 
\begin{pmatrix} 
 a_{11} (t) \mathcal{L}_{n-1} (s,\,k_x) & 0 & a_{13} (t) \mathcal{L}_{n-1} (s,\,k_x) & a_{14} (t)\mathcal{M}_n (s,\,k_x) \\ 
0& 0 & 0 & 0 \\
a_{31} (t) \mathcal{L}_{n-1} (s,\,k_x) & 0 & a_{33}(t) \mathcal{L}_{n-1} (s,\,k_x) & a_{34} (t) \mathcal{M}_{n} (s,\,k_x) \\
 a_{41} (t) \mathcal{M}_n (s,\,k_x) & 0 & a_{43}(t) \mathcal{M}_{n} (s,\,k_x) & a_{44} (t) \mathcal{L}_n (s,\,k_x) \\
\end{pmatrix},
\end{equation} \normalsize
with the time-dependent coefficients given by
\begin{eqnarray}
a_{11} (t) &=& 1 - 4 (A_n ^2 + B_n ^2) \eta ^2 \sin ^2 (E_n t), \label{eqa1} \\
a_{33} (t)&=&- 4 A_n ^2 \eta ^2 \sin ^2 (E_n t), \label{eqa1a} \\
a_{44}(t) &=& - 4 B_n ^2 \eta ^2 \sin ^2 (E_n t), \label{eqa1b} \\
a_{34} (t)&=& a_{43} (t) = - 4 A_n B_n \eta ^2 \sin ^2 (E_n t), \\
a_{13} (t)&=& - a ^* _{31} (t) = - 2 i \eta \sin(E_n t) A_n \big(\cos(E_n t) + i \sin(E_n t)(1 - 2 \eta) \big), \label{eqa2}\\
a_{14} (t) &=& - a ^* _{41} (t) = 2 i \eta \sin(E_n t) B_n \big(\cos(E_n t) + i \sin(E_n t)(1 - 2 \eta) \big). \label{eqa3}
\end{eqnarray}
The phase-space content of the Wigner function is governed by the functions $ \mathcal{L}_{n} (s,\,k_x)$ and $\mathcal{M}_{n} (s,\,k_x)$, given by
\begin{equation}\label{eq46}
\mathcal{L}_{n} (s,\,k_x) = (-1)^n \frac{\sqrt{e \mathcal{B}}}{\pi}\exp[-(s^2 + k_x^2)]L_n[2(s^2 + k_x^2)],
\end{equation}
and 
\begin{equation}\label{eq47}
\mathcal{M}_{n} (s,\,k_x) = \frac{(-1)^n }{2 \pi } \sqrt{\frac{e \mathcal{B}}{n}} \exp[-(s^2 + k_x^2)]\left(\frac{d}{ds} L_n[2(s^2 + k_x^2)] \right),
\end{equation}
where $L_n(z)$ is the $n$-th Laguerre polynomial. These functions form an orthonormal basis with respect to phase-space integrations \cite{Gradshteyn},
\begin{eqnarray}
 \int dx \int dk_x \, \mathcal{L}_{n} (s,\,k_x) &=& 1, \\
 \int dx \int dk_x \, \mathcal{M}_{n} (s,\,k_x) &=& 0,\\
 \int dx \int dk_x \, \mathcal{L}_{n} (s,\,k_x) \mathcal{L}_{m} (s,\,k_x) &=&  \int dx \int dk_x \, \mathcal{M}_{n} (s,\,k_x) \mathcal{M}_{m} (s,\,k_x)\nonumber\\
  &=& \delta_{mn} \frac{\sqrt{e \mathcal{B}}}{2\pi}.\label{eq49}
\end{eqnarray}
These relations suffice to all calculations involving up to the product of two elements of the Wigner matrix. For instance, the normalization is immediately verified,
\begin{equation}
\int dx \int dk_x \, Tr [\omega _{n,1} \gamma_0] = a_{11} - a_{33} - a_{44} = 1, \label{normalizationwignermatrix}
\end{equation}
where the integrand can be regarded as a real, but not necessarily positive, quasi-probability distribution in phase space. Therefore, Eq.~\eqref{normalizationwignermatrix} ensures unitarity of the theory and applies to all acceptable Wigner matrices in the framework of the phase-space quantum mechanics, since it is simply the expression for the conservation of probability. 

\begin{figure}
	\begin{center}
		\includegraphics[scale=0.4]{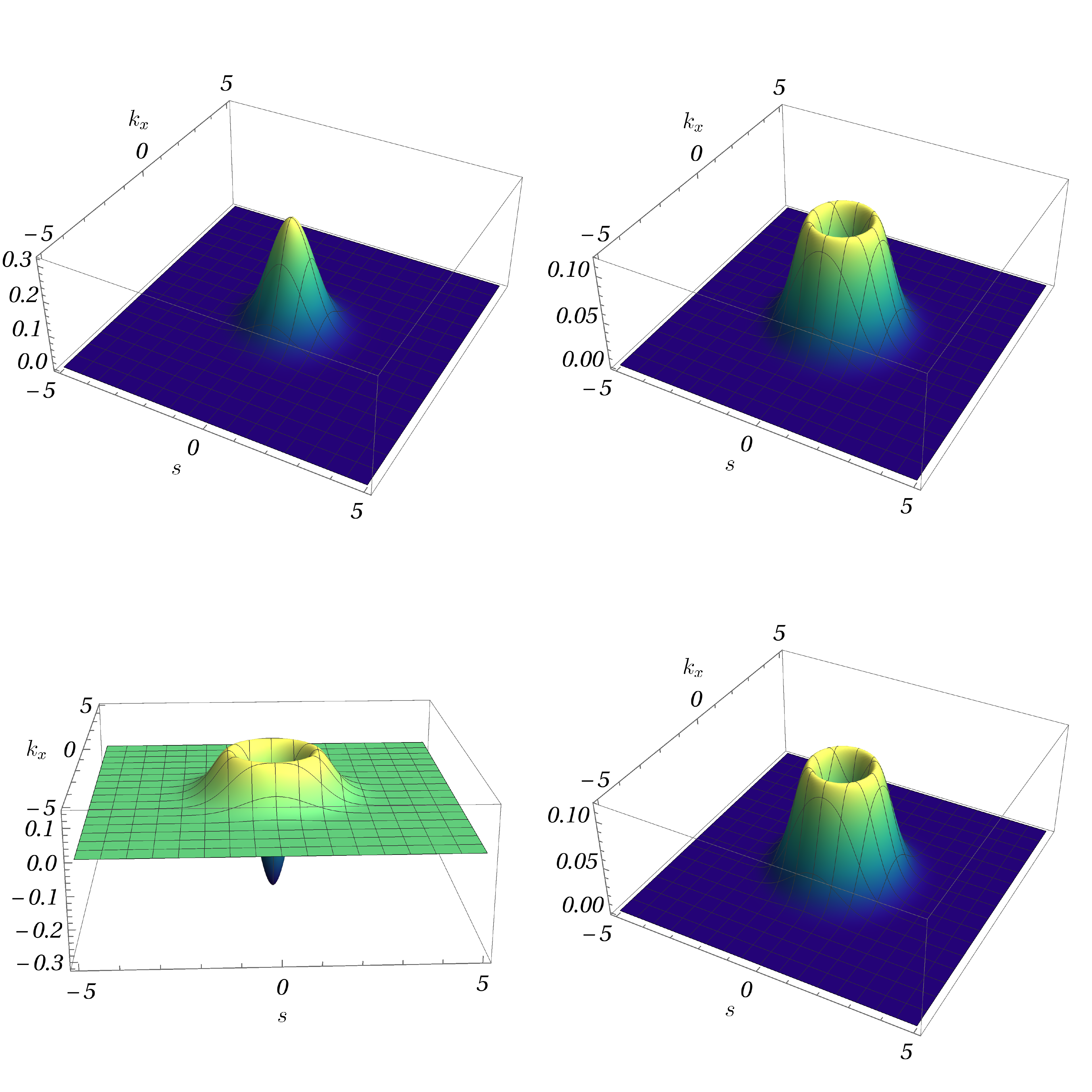}
		\caption[]{Time-evolution of $\frac{1}{\sqrt{e\mathcal{B}}}Tr[\omega_{n=1,i} \gamma^0]$ in phase space $(s,\,k_x)$ for $i=1$ (bottom row) and $i=2$ (top row). States are calculated at $t=0$ (left) and $t= \frac{\pi}{2 E_1}$ (right). The phenomenological parameters have been set to unity, i.e. $k_z = e \mathcal{B} = m = 1$.}
		\label{packet} 
	\end{center}
\end{figure}

Likewise, the averaged behavior in phase space for an initial Gaussian state can be fully described by the Wigner matrix $\omega _{n,1}(s,k_x ;t)$. However, the particular choice of superposition coefficients and eigenstates that contribute to $\mathcal{G}_n ^{^{(i)}} $ fixes not only the initial state polarization, but also the local evolution in phase space (cf. Eq.~\eqref{Gaussianstate1}). More precisely, one could compare the time-evolution of the quasi-probability density as defined by $Tr[\omega_{n,i} \gamma^0]$ for $\mathcal{G}_n ^{^{(1)}} $ and $\mathcal{G}_n ^{^{(2)}} $. The numerical results are shown in Fig.~\eqref{packet}, from which one notices that only $\omega_{n,1} (s, k_x;t)$ corresponds to a Gaussian distribution in phase space for $t=0$ and $n=1$, since $\mathcal{L}_0 (s,\,k_x) \propto \exp[-(s^2 + k_x ^2)]$. As expected, there exists a local spin-parity informational structure within the Wigner function, which coexists with a global (or integrated) one. The correlation profile between these states is indistinguishable upon phase-space averaging, given that the functions $\mathcal{L}_n (s, k_x)$ are orthonormalized. Thus, calculations for averaged properties will be implemented through $\omega_{n,1} (s, k_x;t)$ for convenience.

\begin{figure}
	\begin{center}
		\includegraphics[scale=0.35]{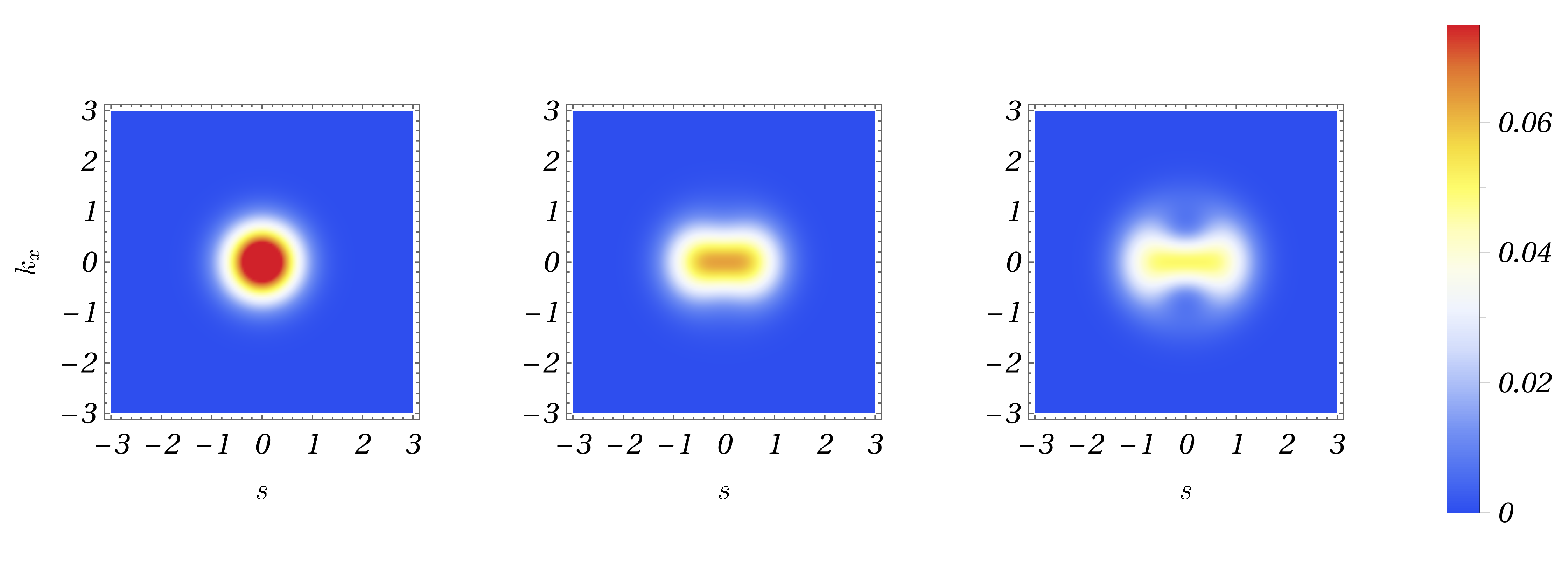}
		\caption[]{Quantum purity, $(1/ e\mathcal{B})Tr [(\omega _{n,1} \gamma_0)^2]$ in phase space $(s,\,k_x)$, evaluated for a Gaussian state ($n=1$). Again, results are for $k_z = e \mathcal{B} = m = 1$, and the color scheme indicates the regions where the state is maximally mixed (blue (dark gray) region). From left to right, $t = (\frac{\pi}{4E_1})j$, for $j=0,\,1,\,2$. The initialized state has a Gaussian quantum purity profile, which spreads along the $s$-direction.}
		\label{figure2}
	\end{center}
\end{figure}

From the quantum purity expression, it is straightforward to check that
\begin{eqnarray}
\mathcal{P} &=& \frac{2\pi}{\sqrt{e \mathcal{B}}} \int dx \int dk_x \, Tr [(\omega _{n,1} \gamma_0)^2] \nonumber\\
&=& a_{11} ^2 + a_{33} ^2 + a_{44}^2 + 2 \vert a_{13}\vert ^2 + 2 \vert a_{34}\vert ^2 + 2 \vert a_{14}\vert ^2, 
\end{eqnarray}
for
\begin{eqnarray}
\vert a_{13} \vert ^2 &=& - a_{11} a_{33}, \\
a_{34} ^2 &=& a_{33} a_{44}, \\
\vert a_{14} \vert ^2 &=& - a_{11} a_{44},
\end{eqnarray}
where the set of orthogonality relations from Eq.~\eqref{eq49} was used.
One can recast the purity expression into the form of
\begin{eqnarray}
\mathcal{P} = ( a_{11} - a_{33} - a_{44})^2 = 1,
\end{eqnarray}
the pure-state constraint. Interestingly, the non-integrated quantum purity spreads preferentially along the $k_x = 0$ due to the contribution from $\mathcal{M}_n ^2 (s,\,k_x) $ for $t\neq0$. Also, from Fig.~\eqref{figure2}, it is possible to see pockets of mixedness (blue (dark gray) regions) surrounded by locally pure regions (white and red regions) as the Wigner function evolves.
Nevertheless, it is worth noticing that the phase-space quantum purity is always a non-negative quantity.

\subsubsection{Classical and quantum correlations}

\begin{figure}
	\begin{center}
		\includegraphics[scale=0.33
		]{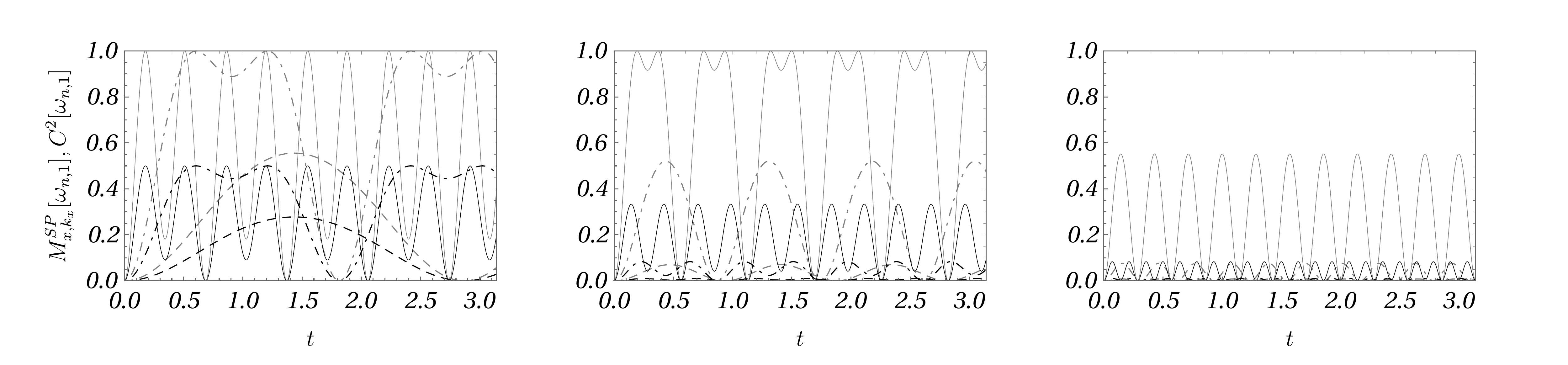}
		\caption[]{Mutual information between phase space and spin-parity space for a Gaussian state (gray lines) and spin-parity quantum concurrence (black lines). For all plots, $m=1$ and, from left to right, $k_z^2 = 0, 10, 100$; one also has $e\mathcal{B} = 1/10, 1,10$ for dashed, dot-dashed, and solid lines, respectively.}
		\label{figure3}
	\end{center}
\end{figure}

From the above result, the mutual information between spin-parity and phase-space degrees of freedom can be assessed for a Gaussian state. It corresponds to the averaged information in phase space that can be inferred from the spin-parity Hilbert space and vice versa. The linear entropy related to spin-parity and phase-space degrees of freedom are
\begin{equation}
\mathcal{I}_{SP} = 1 - Tr \left[ \left(\gamma_0 \langle \omega _{n,1} \rangle \right)^2 \right] = 
8 \sin ^2 (E_n t) \eta_n ^2 B_n ^2 \bigg( 1 - 4B_n ^2 \eta_n ^2 \sin ^2 (E_n t)\bigg),
\end{equation}
and 
\begin{equation}
\mathcal{I}_{\{x,k_x\}} = 1 - \frac{2\pi}{\sqrt{e \mathcal{B}}} \int dx \int dk_x \, \big( Tr \left[\omega _{n,1} \gamma_0 \right]\big) ^2 = 
8 \sin ^2 (E_n t) \eta_n ^2 B_n ^2 \bigg( 1 - 4B_n ^2 \eta_n ^2 \sin ^2 (E_n t)\bigg),
\end{equation}
respectively.
One notices that the linear entropies expressions depend on the quantum number $n$ only through $2n e \mathcal{B}$. Therefore, the Gaussian state $(n=1)$ exhibits the same averaged information profile as a quantum state with arbitrary $n$, given that the factor $2n e \mathcal{B}$ is chosen accordingly. The spin-parity phase-space mutual information (cf. Eq.~\eqref{mut}) reads
\begin{equation} \label{eq52}
M^{SP} _{x, k_x} = 
 16 \sin ^2 (E_n t) \eta_n ^2 B_n ^2 \bigg( 1 - 4B_n ^2 \eta_n ^2 \sin ^2 (E_n t)\bigg),
\end{equation}
which is depicted in Fig.~\eqref{figure3}. 
For $t=0$, the phase-space dependence of the Wigner function is factorized out. Therefore, spin-parity and phase spaces become uncorrelated. The same figure shows that the mutual information oscillates between unity and zero as long as the magnetic field contribution is not suppressed by the $A_1 = k_z/ (E_1 +m)$ coefficient (\ref{parameters}).

It was emphasized that the information measure obtained above amounts to the correlations between spin-parity and phase space coordinates. It is worth noticing, however, that such correlations are not exclusively of quantum nature. Indeed, a quantum state can generally exhibit both, quantum and classical, types of correlation \cite{Henderson}. This assertion can be cleared up in terms of quantum decoherence for two-qubit systems, for which a set of orthogonal projectors are introduced for both contributions related to spin and parity Hilbert spaces so as to be associated to all possible measurements. Then, it is straightforward to check that after any measurement, the off-diagonal elements of the density matrix will unavoidably be damped off in that particular basis \cite{Vedral,Shunlong,Schlosshauer}. Therefore, one can simply consider a stochastic matrix with the probability distribution of the possible outcomes in the standard basis. Thus, considering the localization of the quantum particle, in the basis implied by the Dirac representation, the Wigner function obtained in Eq.~\eqref{wignermatrix} decoheres to a classical-like stochastic matrix, in which the remaining diagonal elements are proportional to probabilities in phase space. Defining such matrix as $\omega_{n,1} ^{\, (cl)} = Diag\,[a_{11} \mathcal{L}_{n-1} \hspace{1em} 0 \hspace{1em} a_{33} \mathcal{L}_{n} \hspace{1em} a_{44} \mathcal{L}_{n-1} ] $, it is possible to observe that, apart from the phase-space coordinate dependence, the elements of the matrix multiplied by $\gamma_0$ are always non-negative. 

In order to quantify the classical contribution to the correlations, the purity computation for $\omega_{n,1} ^{\, (cl)}$ yields
\begin{equation}
\mathcal{P}^{\, (cl)}= a_{11} ^2 + a_{33} ^2 + a_{44} ^2 \leq 1,
\end{equation}
where the equality holds only for $t=0$ and thus confirms that the decohered state is not a pure state, reflecting the loss of information upon measurement. The relative linear entropies are also calculated,
\begin{equation}
\mathcal{I}^{\, (cl)} _{SP} = 1- a_{11} ^2 - a_{33} ^2 - a_{44} ^2, 
\end{equation}
and 
\begin{equation}
\mathcal{I}^{\, (cl)}_{\{x,k_x\}} = 1- a_{11} ^2 - a_{33} ^2 - a_{44} ^2 + 2 a_{11} a_{33},
\end{equation}
which explicitly yields the mutual information (\ref{mut}) between spin-parity and phase spaces for the decohered Wigner function $\omega_{n,1} ^{\, (cl)} $,
\begin{equation}\label{eq56}
M^{SP} _{x, k_x} = -32 B_n ^4 \eta_n ^4 \sin ^4(E_n t) + 8 B_n ^2 \eta_n ^2 \sin ^2 (E_n t) + 32 B_n ^2 A_n ^2 \eta_n ^4 \sin ^4 (E_n t).
\end{equation}
This result shows that there is indeed a certain amount of correlation between the continuous and discrete degrees of freedom that is of classical-like nature. Therefore, the difference between the total mutual information and the above expression results into a correlation of quantum nature. For pure states, quantum correlation implies into entanglement \cite{Shunlong}, which is quantified by the so-called quantum concurrence. As a matter of fact, by computing the quantum concurrence for the Wigner function, it will be shown that indeed the spin-parity non-separability codified by the Wigner function is regarded as the quantum-like information on the Hilbert space associated to the continuous degrees of freedom that can be inferred from the spin-parity space.

For a pair of qubits, concurrence is a well-defined entanglement measure, which in turn is related to the more physically appealing entanglement of formation (EoF). More precisely, for pure states,\footnote{Otherwise, it is defined as the average entanglement of the pure states that realize the given density matrix, minimized over all decompositions on pure states. } EoF is monotonically increasing for $0 \leq C \leq 1$, which is always the case. It is defined by \cite{n024}
\begin{eqnarray}
E_{EoF} [\varrho] &=& \mathcal{E}\left[ \frac{1 - \sqrt{1 - \mathcal{C}^2[\varrho]}}{2}\right],
\end{eqnarray}
with $\mathcal{E}[\lambda] =- \lambda \log_2 \lambda - (1-\lambda)\log_2 (1-\lambda)$ and the quantum concurrence defined as
\begin{equation}\label{condef}
\mathcal{C}[\varrho] = \sqrt{\langle w \vert \widetilde{\rho}\vert w\rangle} = \vert \langle w \vert \widetilde{w} \rangle\vert = \sqrt{Tr [ \varrho \widetilde{\varrho}]},
\end{equation}
for a pure state $ \varrho = \vert w \rangle\langle w\vert$, where $\vert \widetilde{w} \rangle $ the spin-flipped state, 
\begin{equation}
\vert \widetilde{w} \rangle = \sigma^{(1)}_y\otimes\sigma^{(2)}_y\vert {w}^* \rangle,
\end{equation}
with ``$*$'' denoting the complex conjugation operator.

Once the identification of the density matrix for a pair of qubits with the matrix-valued Wigner function is made, quantum concurrence can be computed in a straightforward fashion. In order to describe the phase-space pattern of the quantum concurrence, the density matrix is identified as  $\varrho \equiv \gamma^0 \, \omega _{n,1} $ and from the $SU(2) \otimes SU(2)$ decomposition of $\gamma^2 = i \sigma^{(1)}_y\otimes\sigma^{(2)}_y$  \cite{BernardiniEPJP}, it follows that the spin-flipped density matrix is identified as $\widetilde{\varrho} \equiv \,(-i\gamma^{2})\gamma^{0}\,\omega ^* _{n,1} \,(-i\gamma^{2}) $ for any Wigner function under consideration \eqref{wignermatrix}. Then, the local quantum concurrence reads
\begin{eqnarray}
\mathcal{C}^2[\omega_{n,1}] (s,\,k_x) &=& (-1) Tr[\omega_{n,1} \,\gamma^2\gamma^{0}\,\omega^* _{n,1} \,\gamma^2\gamma^{0}] \nonumber \\
&=& 8 \eta_n ^2 \sin ^2 (E_n t) B_n ^2 \big[1-4(A_n ^2 + B_n^2) \eta_n ^2 \sin ^2 (E_n t)\big]\nonumber \\
&& \hspace{6em} \times \, \big( \mathcal{L}_n (s,\,k_x) \mathcal{L}_{n-1} (s,\,k_x) + \mathcal{M}^2 _n (s,\,k_x) \big).
\end{eqnarray}
\begin{figure}
	\includegraphics[scale=0.3
	]{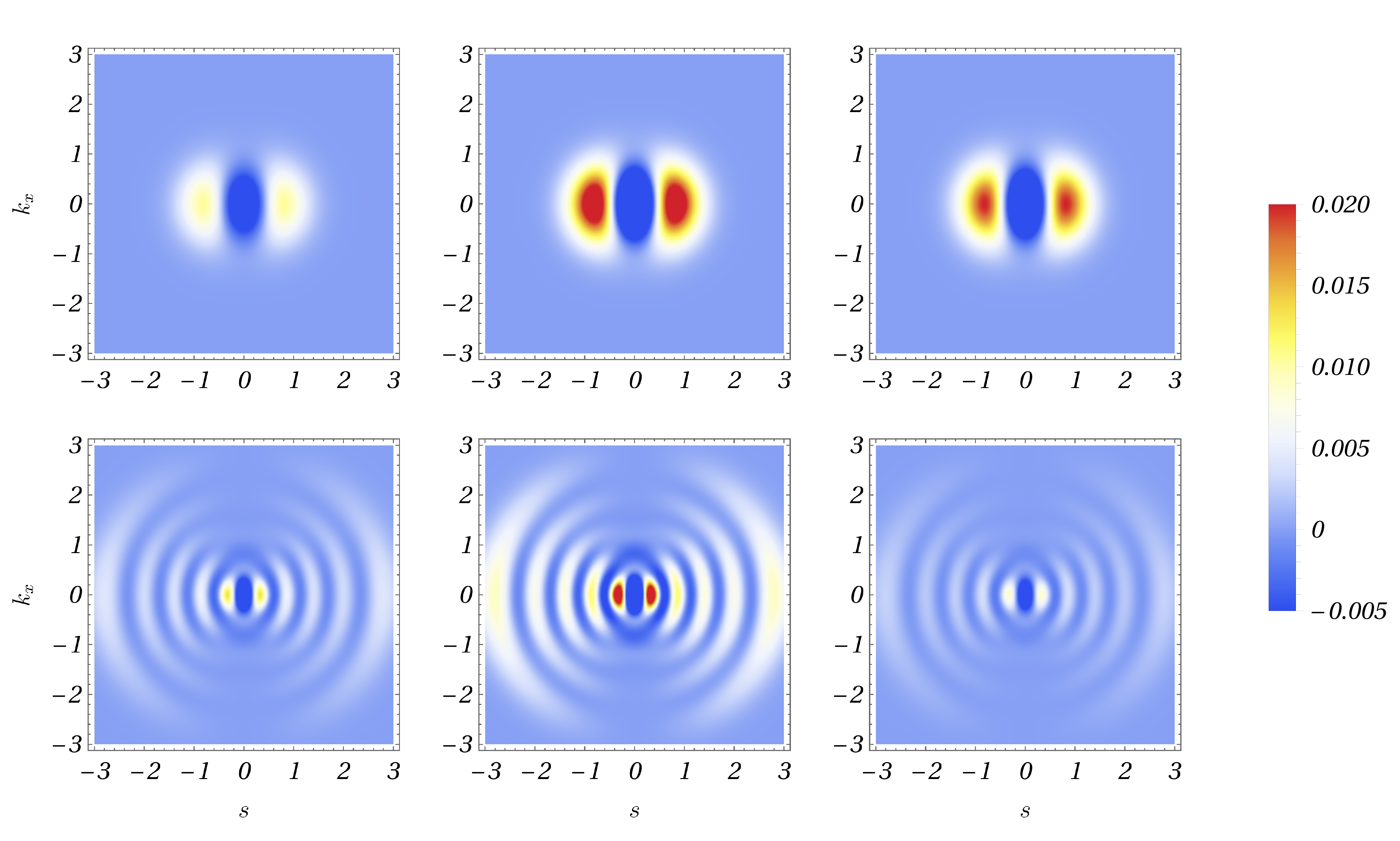}
	\caption[]{Time-evolution of the phase-space dependent spin-parity quantum concurrence $(1/ e\mathcal{B})\mathcal{C}^2[\omega_{n,1}] (s,\,k_x)$ for a departing Gaussian state (top row), with $n=1$, and for $n=5$ (bottom row). From left to right, $t = (\pi/j)/E_n$ for $j = 8,\, 4,\, 2$. Additional parameters follow Fig.~\eqref{figure2}. It is possible to see that the local profile of the quantum concurrence is not positively defined due to the intrinsic correlation with the continuous degrees of freedom themselves.}
	\label{localconc}
\end{figure} 
The product $\mathcal{L}_n (s,\,k_x) \mathcal{L}_{n-1} (s,\,k_x) $ implies that the phase-space profile can exhibit regions of negativity due to the correlation between spin-parity and phase-space degrees of freedom, which is depicted in Fig.~\eqref{localconc}. After averaging over phase space coordinates, the above expression yields the spin-parity non-separability as 
\begin{equation}\label{concurrence2}
\mathcal{C}^2_{SP} = 
8 \eta_n ^4 \sin ^2 (E_n t)B_n^2 \left(\frac{1}{\eta_n ^2} - 4 \sin ^2 (E_n t)(A_n ^2 + B_n ^2) \right),
\end{equation} 
which is the proper quantum concurrence measure for a Gaussian state. The same expression would have been obtained had one considered $\varrho \equiv \gamma^0 \, \langle \omega _{n,1} \rangle$, that is, when the phase-space degrees of freedom are averaged out before the computation of the quantum concurrence. Therefore, only the discrete degrees of freedom are relevant in this computation. 

What stands out in this result is that the quantum concurrence squared indeed corresponds to the difference between the (total) correlations between spin-parity and phase-space degrees of freedom from Eq.~\eqref{eq52} and the classical correlations from Eq.~\eqref{eq56}, as it was previously advertised in Fig.~\eqref{figure3}. Therefore, quantum concurrence is then regarded as a strictly quantum correlation measure such that separable states are easily identified from a particular choice of parameters. For $t = (l \pi)/E_n$, with $l$ integer, there corresponds the initial state which is indeed separable. In the massless limit, i.e. $A_n ^2 + B_n ^2 = 1$ (\ref{parameters}), concurrence vanishes for $t = \pi(l + 1/2)/E_n $ ($l$ integer). 

Just for completeness, concurrence can also be related to the phenomenon of chiral oscillation \cite{extfields}. It has been shown that the averaged values of the chiral operator $\hat{\gamma}_5$ coincide with the critical points of the concurrence for constant external potentials \cite{diraclike01}. Here, chiral projections are obtained from the Wigner matrix itself, $\omega_{L,R} = P_{L,R} \, \omega $, with the usual left and right projectors, $P_L = (1- \hat{\gamma}_5)/2 $ and $P_R = (1+ \hat{\gamma}_5)/2 $. Since $P_{L,R} P_{L,R} = P _{L,R}$ and $P_{L,R} P_{R,L} = 0$, a chiral projection exhibits no quantum concurrence. This suggests that concurrence can be affected by the interference between chiral projections. To check this, one can evaluate the average chirality with
\begin{eqnarray}
\langle \gamma_5 \rangle &=& \int dx \int dk_x \, Tr \left[ \omega _{n,1} \gamma_0 \gamma_5 \right] \nonumber \\
&=& \frac{4 \eta A_n m}{E_n} \sin ^2(E_n t), 
\end{eqnarray}
which is constrained to $0 \leq \langle \gamma_5 \rangle \leq 1$, a non-negative chirality due to the choice of the particular polarization of $\mathcal{G}_{n} ^{^{(1)}} (s,\,t)$.
\begin{figure}
	\begin{center}
		\includegraphics[scale=0.4
		]{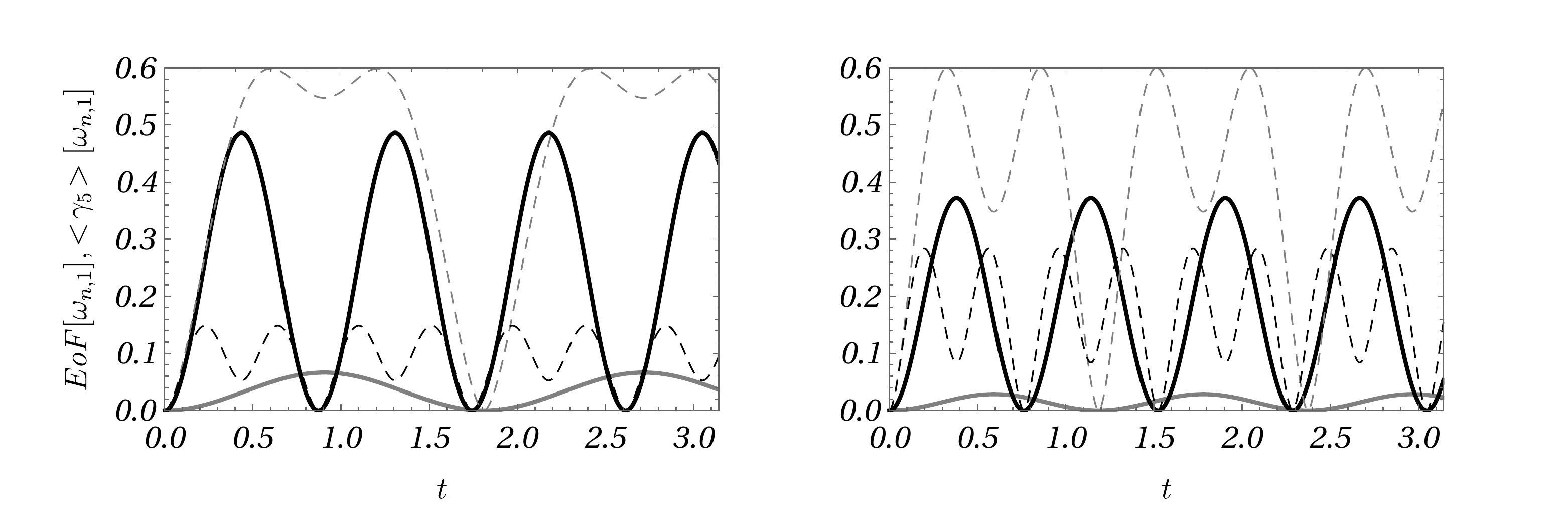}
		\caption[]{$EoF$ (dashed lines) and chiral oscillation in terms of $\langle \gamma_5 \rangle (t)$ for a Gaussian state. One has $e\mathcal{B} = 1,\,3$ from left to right and $k_z^2 = 1/100,\,10$ for gray and black lines, respectively, for unity mass.}\label{figure5}
	\end{center}
\end{figure} 
More relevantly, since the amplitude of the averaged chiral oscillation is proportional to $2k_z m/E_n ^2$, it is suppressed for stronger magnetic fields, whereas the concurrence oscillation grows. For instance, the greatest value of $\vert \langle \gamma_5 \rangle \vert = 1 $ would only be obtained for $B_n = \sqrt{2n e \mathcal{B}}/(E_n + m) = 0$, for which the state is separable. Moreover, when $\vert \langle \gamma_5 \rangle \vert$ is at a local maximum, quantum concurrence is at a local minimum. This behavior is depicted in terms of the $EoF$ in Fig.~\eqref{figure5}.

To partially summarize, the dynamics of the local and global information profile has been analyzed for a quantum fermion prepared as a Gaussian state. It was shown that the mutual information between discrete and continuous degrees of freedom encompasses both classical and quantum correlations; moreover, the latter exhibits a close connection to chiral oscillation, due to the fact that chiral projections are spin-parity separable Wigner functions. 

In the next subsection, the theoretical tools for calculating the phase-space averaged information profile shall be extended to the cat state configurations introduced by Eq.~\eqref{SCS}. In order to do so, the phase-space dependence of the Wigner matrix, previously given by Laguerre polynomials (and their first derivatives), will be described in terms of the generalized Laguerre polynomials, since the Dirac cat states in phase space involve two arbitrary principal quantum numbers. 
 
\subsection{Wigner matrix for Dirac cat states}

Considering the information profile for a Gaussian state centered at the origin discussed previously, one should inquire on possible generalizations; namely, how correlations between spin-parity and phase spaces are affected by superposing two Gaussian states at arbitrary distances from the origin.

A symmetric superposition of Gaussian states will be considered, for which the matrix obtained with the standard matrix multiplication is $\phi^S \bar{\phi}^S$ (cf. Eq.~\eqref{SCS}). In order to compute the correspondent Wigner function for $\phi^S \bar{\phi}^S$, the $1$-dim spatial intrinsic integral from the Weyl transform will be expressed by terms such as 

\begin{equation}\label{wignerhermite}
\pi^{-1} \int du \, e^{2iku} e^{-(s+u)^2/2}e^{-(s-u)^2/2} H_{n} (s-u) H_{m} (s+u)
\end{equation}
for $n,\, m =0,\,1,\,2...$ accounting for all terms in the infinite series from Eqs.~(\ref{qstte1})-(\ref{qstte3}). One then needs to consider $n\geq m$ and $n \leq m$ separately \cite{Gradshteyn}, which leads to
\begin{equation}
(-2)^n \pi^{-1/2} (m!) \exp[-(s^2+k^2)] (-s + ik)^{n -m} L^{n-m} _m (2(k^2 + s^2)), 
\end{equation}
for $n \geq m$, and to
\begin{equation}
(-2)^n \pi^{-1/2} (n!) \exp[-(s^2+k^2)] (+s + ik)^{m -n} L^{m-n} _n (2(k^2 + s^2)).
\end{equation}
for $m \geq n$. The functions $L^{l_1} _{l_2} (z^2)$ are the generalized Laguerre polynomials at the phase-space radius $z^2 = 2(k^2 + s^2)$ and only occur here with natural indices \cite{Weniger}. Of course, for $n=m$, both expressions concur. 

When these expressions appear in summations, it will be helpful to implement orthogonality relations in phase space. By collecting the factors from the normalized function $\mathcal{F}_{n} (s)$, one defines
\begin{eqnarray}\small
\mathfrak{L}_{mn}=\begin{cases}
\left(\mathcal{L}_{m} ^{ ^{( n-m) }}\right)^* = \frac{\sqrt{e \mathcal{B}}}{\pi} \left(\frac{m!}{n!}\right)^{1/2} (-1)^m e^{-(s^2 +k_x ^2)} [2^{1/2}(s - i k_x)]^{n-m} L_{m} ^{^{(n-m)}} [2(k_x ^2 + s^2)], n \geq m, \\
 \mathcal{L}_{n} ^{^{( m-n) }}= \frac{\sqrt{e \mathcal{B}}}{\pi} \left(\frac{n!}{m!}\right)^{1/2}(-1)^n e^{-(s^2 +k_x ^2)} [2^{1/2}(s + i k_x)]^{m-n} L_{n} ^{^{(m-n)}} [2(k_x ^2 + s^2)], m \geq n,
\end{cases}\label{normalizedfunction}
\end{eqnarray}\normalsize
where the phase-space dependence was omitted on the left-hand side for clarity of notation. The notation $(...)^*$ was introduced to indicate complex conjugation followed by a swapping of indices. The function components $\mathfrak{L}_{mn}$ satisfy \cite{Weniger,Gradshteyn}
\begin{equation}\label{eq75}
\int dx \int dk_x \, \mathfrak{L}_{mn} = \delta_{mn},
\end{equation}
and
\begin{eqnarray}\label{eq76}
\int dx \int dk_x \, \mathfrak{L}_{mn} \mathfrak{L}_{m'n'} = 
\frac{\sqrt{e \mathcal{B}}}{2\pi}\delta_{m n'} \delta_{n m'}, 
\end{eqnarray}
which compose the relations associated to normalization and purity conditions of the Wigner matrix. The double integrals in Eqs.~(\ref{eq75})-(\ref{eq76}) are evaluated as in standard integration of Laguerre-type functions in polar coordinates and going to the complex plane in the $k_x$ variable \cite{Gradshteyn}. Fortunately, they suffice to calculate all quantities related to the averaged correlation profile between spin and parity Hilbert spaces. 

Once the phase-space structure is settled, all elements of the Wigner matrix for cat states can be readily obtained. For instance,
\begin{eqnarray}\label{wigner11}
\mathcal{W}_{11} (s,k_x;t) = \mathcal{N}_a\sum_{ \{m,n \} \text{odd} } \frac{\bigg( e^{i E_{n}t} + (A^{2} _{n} + B^{2} _{n}) e^{-i E_{n}t} \bigg)}{1+ A^{2} _{n} + B^{2} _{n} } && \hspace{-0.75em} \frac{ \bigg( e^{-i E_{m}t} + (A^{2} _{m} + B^{2} _{m}) e^{i E_{m}t} \bigg)}{1+ A^{2} _{m} + B^{2} _{m} } \times \quad\nonumber \\
&& \frac{(a/\sqrt{2})^{n+m-2}}{\sqrt{\Gamma(n)\Gamma(m)}} \mathfrak{L}_{_{(m-1) (n-1)}}(s,\,k_x),
\end{eqnarray}
where $S$-cat index has been omitted, $\Gamma(n)=(n-1)!$ is the gamma function, $\mathcal{N}_a$ is the normalization constant to be determined and, for convenience, the indices run over odd numbers. The time-dependent factor of the $n$-th term comes from $\phi ^\dagger (s,\,t)$ (cf. Eq.~\eqref{SCS}), whereas the $m$-th term comes from $\phi (s,\,t)$. The remaining diagonal terms of the Wigner matrix are thus
\begin{eqnarray}
\mathcal{W}_{33} (s,k_x;t) &=& -4\mathcal{N}_a \hspace{-0.75em} \sum_{ \{m,n \} \text{odd} } \frac{\sin(E_{n}t) A_{n}}{1+ A^{2} _{n} + B^{2} _{n} } \frac{ \sin(E_{m}t) A_{m}}{1+ A^{2} _{m} + B^{2} _{m} } \frac{(a/\sqrt{2})^{n+m-2}}{\sqrt{\Gamma(n)\Gamma(m)}} \mathfrak{L}_{_{(m-1)(n-1)}} (s,\,k_x),\,\,\, \label{wigner33}\\\label{wigner44}
\mathcal{W}_{44} (s,k_x;t) &=& -4\mathcal{N}_a \hspace{-0.75em} \sum_{ \{m,n \} \text{odd} } \frac{\sin(E_{n}t) B_{n}}{1+ A^{2} _{n} + B^{2} _{n} } \frac{ \sin(E_{m}t) B_{m}}{1+ A^{2} _{m} + B^{2} _{m} } \frac{(a/\sqrt{2})^{n+m-2}}{\sqrt{\Gamma(n)\Gamma(m)}} \mathfrak{L}_{mn} (s,\,k_x).
\end{eqnarray}
All diagonal terms are real-valued, as it should be, and the non-diagonal elements are given by
\small
\begin{eqnarray}
\mathcal{W}_{31} (s,k_x;t) &=& - 2i\mathcal{N}_a \hspace{-1em} \sum_{ \{m,n \} \text{odd} } \hspace{-1em} \frac{\bigg( e^{i E_{n}t} + (A^{2} _{n} + B^{2} _{n}) e^{-i E_{n}t} \bigg)}{1+ A^{2} _{n} + B^{2} _{n} } \frac{ \sin(E_{m}t) A_{m}}{1+ A^{2} _{m} + B^{2} _{m} } \frac{(a/\sqrt{2})^{n+m-2}}{\sqrt{\Gamma(n)\Gamma(m)}} \mathfrak{L}_{_{(m-1)(n-1)}} (s,\,k_x) \nonumber \\
&=&- \mathcal{W} ^* _{13} (s,k_x;t), \\
\mathcal{W}_{41} (s,k_x;t) &=& 2i\mathcal{N}_a \hspace{-1em} \sum_{ \{m,n \} \text{odd} } \hspace{-1em} \frac{\bigg( e^{i E_{n}t} + (A^{2} _{n} + B^{2} _{n}) e^{-i E_{n}t} \bigg)}{1+ A^{2} _{n} + B^{2} _{n} } \frac{ \sin(E_{m}t) B_{m}}{1+ A^{2} _{m} + B^{2} _{m} } \frac{(a/\sqrt{2})^{n+m-2}}{\sqrt{\Gamma(n)\Gamma(m)}} \mathfrak{L}_{_{(m)(n-1)}} (s,\,k_x) \label{wigner41} \nonumber \\
&=& - \mathcal{W} ^* _{14} (s,k_x;t), \\
\mathcal{W}_{34} (s,k_x;t) &=& - 4\mathcal{N}_a \hspace{-1em} \sum_{ \{m,n \} \text{odd} } \frac{ \sin(E_{n}t) A_{n} }{1+ A^{2} _{n} + B^{2} _{n} } \frac{ \sin(E_{m}t) B_{m}}{1+ A^{2} _{m} + B^{2} _{m} } \frac{(a/\sqrt{2})^{n+m-2}}{\sqrt{\Gamma(n)\Gamma(m)}} \mathfrak{L}_{_{(m)(n-1)}} (s,\,k_x) \nonumber \\
&=& \mathcal{W} ^* _{43} (s,k_x;t). \label{wigner43}
\end{eqnarray}\normalsize
\begin{figure}
	\includegraphics[scale=0.3]{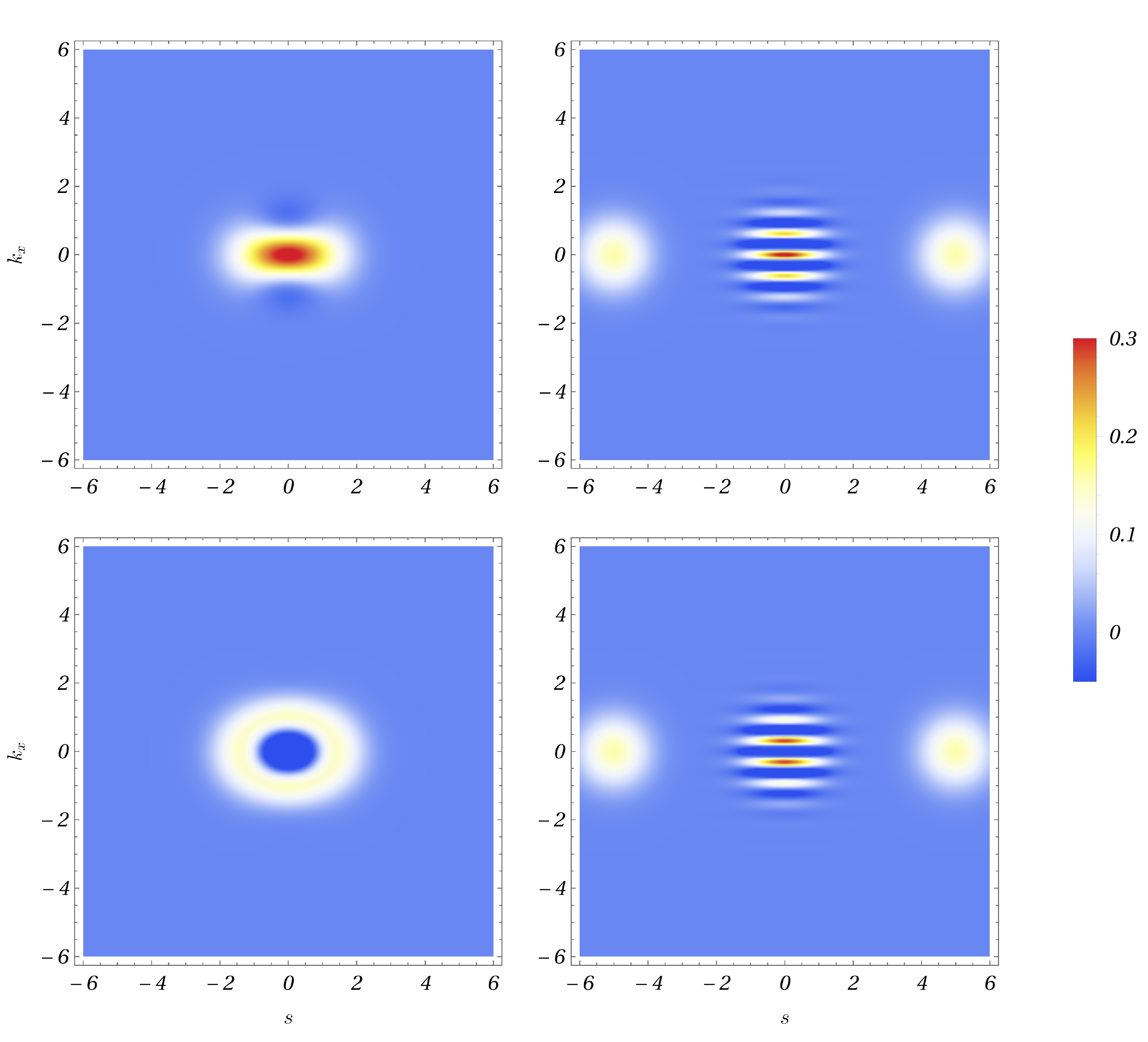}
	\caption[]{Phase-space $(s,\,k_x)$ quasi-probability density for Dirac cat states, $\frac{1}{\sqrt{e \mathcal{B}}}Tr[\mathcal{W}(s,k_x;t)\,\gamma_0]$, centered at $a=\pm 1$ (left column) and $a= \pm 5$ (right column). Both symmetrical (top) and anti-symmetric (bottom) superpositions are displayed with corresponding parameters $k_z^2 = e\mathcal{B} = 1$. For increasing values of $a$, their phase-space profile becomes barely distinguishable, whereas the overlapping of Gaussian states as $a \rightarrow 0$ shows that the amplitude for the $A$-state is suppressed.}
\label{figure6}
\end{figure}
From the orthogonality relations of $ \mathfrak{L}_{mn} (s,\,k_x)$, the elements $\mathcal{W}_{\mu \nu}$ with $\mu + \nu$ odd always integrate out to zero in phase space, since they contain terms in the form $ \mathfrak{L}_{_{(m)(n-1)}}$, i.e. an even-odd combination. Nevertheless, quadratic terms generally do not average out to zero; thus, they can be regarded as a generalization of the functions for a definite quantum number found in the previous subsection. To clear up this assertion, one considers, for instance, the sum of functions $ \mathfrak{L}_{mn} (s,\,k_x)$ whose indices differ by unity, $\mathfrak{L}_{(l)(l+1)} + \mathfrak{L}_{(l+1)(l)}$,
\begin{eqnarray}
\left(\mathcal{L}_{l} ^{^{(1) }}\right)^* + \mathcal{L}_{l} ^{^{(1) }}&=& \frac{2^{3/2}(l+1)^{-1/2}}{\pi} s \exp[-(s^2 + k^2)] L_l ^{^{(1)}} (2(k^2 + s^2)) \nonumber \\
&=& 2 \mathcal{M}_{l+1} (s,\,k_x),
\end{eqnarray}
where $\mathcal{M}_{l+1} (s,\,k_x) $ was obtained previously in Eq.~\eqref{eq47}. On the other hand, for $n=m$,
\begin{equation}
\mathcal{L}_{n} ^{^{(0) }} = \left(\mathcal{L}_{n} ^{^{(0) }}\right)^* = (-1)^n \frac{\sqrt{e \mathcal{B}}}{\pi}\exp[-(s^2 + k_x^2)]L_n[2(s^2 + k_x^2)],
\end{equation}
which is the function $\mathcal{L}_n (s,\,k_x)$ from Eq.~\eqref{eq46}. Therefore, given that cat states are a superposition of Gaussian states in configuration space, it is possible to identify the corresponding superposition law in phase space as well.

Even though the Wigner matrix is expressed by several combinations of infinite series expansions, many properties can be analytically replicated. For instance, the normalization is calculated with
\begin{eqnarray}\small \label{normaizationspinor}
&&\int dx \int dk_x \, \mbox{Tr}[\mathcal{W}(s,k_x;t)\,\gamma_0] = \int dx \int dk_x \, (\mathcal{W}_{11} - \mathcal{W}_{33} - \mathcal{W}_{44}) \nonumber \\
&=&\mathcal{N}_a\sum_{n \, \text{odd}} \eta _{n} ^2 \big\{ 1 +( A_{n} ^2 + B_{n} ^2)^2 + 2( A_{n} ^2 + B_{n} ^2)\cos(2E_{n}t) - 4\sin ^2(E_{n}t)( A_{n} ^2 + B_{n} ^2) \big\} \frac{(a^2/2)^{n-1}}{{(n-1)!}} \nonumber \\
&=& \mathcal{N}_a\,\cosh(a^2/2)=1, 
\end{eqnarray}\normalsize
where the integrals are evaluated in terms of the orthonormalization conditions from \eqref{eq75}.

Even if the above results were concerned with the $S$-states superposition, $A$-state superposition can be equivalently evaluated. In this case, the series expansions from Eqs.~\eqref{wigner11}-\eqref{wigner43} shall have their indices running over even (replacing odd) numbers only. This is in agreement with the fact that for $t=0$ the only non-vanishing element of the Wigner function has phase-space functions $\mathfrak{L}_{(m-1)(n-1)}$ with $m-1$ and $n-1$ odd, so all algebraic manipulations remain valid, except for the normalization constant $\mathcal{N}_a=\cosh(a^2/2)^{-1}$ (cf. \eqref{normaizationspinor}) which is replaced by  $\mathcal{N}_a=\sinh(a^2/2)^{-1}$. If $t=0$, the phase-space pattern of Eq.~\eqref{normaizationspinor} can be indirectly obtained by the computation of the more straightforward Wigner function from Eq.~\eqref{catstate}; that is, the cat state in configuration space. Similarly, the $A$-state is also depicted in Fig.~\eqref{figure6}. 
 
From now on, the normalized Wigner function is implied by multiplying $\mathcal{W}_{\mu\nu}$ either by $\cosh ^{-1}(a^2/2) $ (for the symmetric state) or by $\sinh ^{-1}(a^2/2)$ (for the anti-symmetric state) in order to have the unitarity preserved. In this way, the quantum informational aspects of cat states can be finally assessed. 

Moving to the computation of the relative linear entropies and quantum purity, one notices that the manipulation of the infinite series can be quite intricate; nevertheless, mathematical identities often dispense with the actual computation of the whole expression. To see this, with $\mathcal{{W}}_{\mu \nu}(s,\,k_x;\,t) \equiv \mathcal{W}_{\mu \nu}$ for compactness of notation, one then has
\begin{equation}
\mathcal{P} = \frac{2\pi}{\sqrt{e \mathcal{B}}} \bigg( \langle \mathcal{W}_{11}^2\rangle + \langle \mathcal{W}_{33} ^2 \rangle + \langle \mathcal{W}_{44} ^2 \rangle - 2 \langle \mathcal{W}_{13}\mathcal{W}_{31} \rangle + 2\langle \mathcal{W}_{34}\mathcal{W}_{43} \rangle -2 \langle \mathcal{W}_{14}\mathcal{W}_{41} \rangle \bigg),
\end{equation}
for the purity expression.
Although it might seem intractable, each term can be re-written as
\begin{eqnarray}
\langle \mathcal{W}_{11} \rangle ^2 & =& \cosh(a^2/2) ^{-2} \left( \sum_{n \text{ odd} } \eta_n ^2 \bigg | e^{-i E_nt} + (A^{2} _n + B^{2} _n ) e^{i E_n t} \bigg |^2 \frac{(a/\sqrt{2})^{2n-2}}{\Gamma(n)} \right)^2 \nonumber \\
&=& \cosh(a^2/2) ^{-2} \left( \cosh(a^2/2) - 4 \sum_{n \text{ odd} } \eta_n ^2\sin ^2 (E_n t) (A^{2} _n + B^{2} _n ) \frac{(a^2/2)^{n-1}}{\Gamma(n)} \right)^2 \nonumber \\
&=& \left( 1 - \frac{4}{ \cosh(a^2/2) } \sum_{n \text{ odd} } \eta_n ^2\sin ^2 (E_n t) (A^{2} _n + B^{2} _n ) \frac{(a^2/2)^{n-1}}{\Gamma(n)} \right)^2 \nonumber \\
&=& \frac{2\pi}{\sqrt{e \mathcal{B}}} \langle \mathcal{W}_{11} ^2 \rangle, \label{eq79}
\end{eqnarray}
where the last equality is obtained by noticing that, upon integration of $\mathcal{W}_{11}^2$, one can use the set of relations from Eqs.~(\ref{eq75})-(\ref{eq76}). The explicit calculation is presented in Appendix \ref{appendix}. Similarly,
\begin{eqnarray}
\langle \mathcal{W}_{33} \rangle ^2 = \frac{2\pi}{\sqrt{e \mathcal{B}}} \langle \mathcal{W}_{33} ^2 \rangle & =& 16 \cosh(a^2/2) ^{-2} \left( \sum_{n \text{ odd} } \eta_n ^2 \sin ^2(E_n t) A_n ^2 \frac{(a^2/2)^{n-1}}{\Gamma(n)} \right)^2, \label{eq80}\\
\langle \mathcal{W}_{44} \rangle ^2 = \frac{2\pi}{\sqrt{e \mathcal{B}}} \langle \mathcal{W}_{44} ^2 \rangle & =& 16 \cosh(a^2/2) ^{-2} \left( \sum_{n \text{ odd} } \eta_n ^2 \sin ^2(E_n t) B_n ^2 \frac{(a^2/2)^{n-1}}{\Gamma(n)} \right)^2, \label{eq81}\\
\langle \mathcal{W}_{13}\mathcal{W}_{31} \rangle &=& \langle \mathcal{W}_{11} ^2 \rangle ^{1/2} \langle \mathcal{W}_{33} ^2 \rangle ^{1/2}, \\
\langle \mathcal{W}_{34}\mathcal{W}_{43} \rangle &=& \langle \mathcal{W}_{33} ^2 \rangle ^{1/2} \langle \mathcal{W}_{44} ^2 \rangle ^{1/2}, \\
\langle \mathcal{W}_{14}\mathcal{W}_{41} \rangle &=& \langle \mathcal{W}_{11} ^2 \rangle ^{1/2} \langle \mathcal{W}_{44} ^2 \rangle ^{1/2}.
\end{eqnarray}
The purity expression finally results into
\begin{eqnarray}
\mathcal{P} &= & \bigg( \langle \mathcal{W}_{11} \rangle - \langle \mathcal{W}_{33} \rangle - \langle \mathcal{W}_{44} \rangle \bigg)^2 \nonumber \\
&=& 1,
\end{eqnarray}
where the expression inside the brackets yields the normalization condition as calculated in Eq.~\eqref{normaizationspinor}. Thus, the cat states indeed correspond to a pure state. 

The spin-parity relative entropy is similarly calculated with
\begin{equation}
\mathcal{I}_{SP} = 1 - \langle \mathcal{W}_{11} \rangle^2 - \langle \mathcal{W}_{33} \rangle ^2 - \langle \mathcal{W}_{44} \rangle ^2 + 2 \langle \mathcal{W}_{13}\rangle \langle \mathcal{W}_{31} \rangle,
\end{equation}
of which only the last term needs to be evaluated. $\mathcal{W}_{13}$ is the only non-diagonal element that does not average out to zero. Instead, one can easily verify that
\begin{eqnarray}
\langle \mathcal{W}_{31} \rangle & = & - 2i \cosh(a^2/2) ^{-1} \sum_{n \text{ odd} } \eta_n ^2 \left( e^{-i E_nt} + (A^{2} _n + B^{2} _n ) e^{i E_n t} \right) \sin(E_n t) A_n \frac{(a^2/2)^{n-1}}{\Gamma(n)},
\end{eqnarray}
a complex-valued expression. However, as expected, 
\begin{eqnarray}\label{eq3113}
\langle \mathcal{W}_{31} \rangle \langle \mathcal{W}_{13} \rangle = 4 \bigg | \cosh(a^2/2) ^{-1} \hspace{-0.3em} \sum_{n \text{ odd} } \hspace{-0.3em} \eta_n ^2 \hspace{-0.2em} \left( e^{-i E_nt} + (A^{2} _n + B^{2} _n ) e^{i E_n t} \right) \sin(E_n t) A_n \frac{(a^2/2)^{n-1}}{\Gamma(n)} \bigg | ^2
\end{eqnarray}
is real. For the position-momentum relative entropy, one has
\begin{equation}
\mathcal{I}_{\{x,k_x\}} = 1 - \frac{2\pi}{\sqrt{e\mathcal{B}}} \bigg(\langle \mathcal{W}_{11} ^2 \rangle - \langle \mathcal{W}_{33} ^2 \rangle - \langle \mathcal{W}_{44} ^2 \rangle + 2 \langle \mathcal{W}_{11} \mathcal{W}_{33} \rangle \bigg),
\end{equation}
where it has been used that $\langle \mathcal{W}_{11} \mathcal{W}_{44} \rangle = \langle \mathcal{W}_{33} \mathcal{W}_{44} \rangle = 0$.\footnote{Since one needs to evaluate integrals of products such as $ \mathfrak{L}_{(m-1)(n-1)} (s,\,k_x) \mathfrak{L}_{m'n'} (s,\,k_x)$ with dummy indices being odd.} Only the right-most term was not calculated yet, and the computation follows along the same lines of the previous identities (cf. Appendix \ref{appendix}). Then, 
\begin{equation}
\frac{2\pi}{\sqrt{e\mathcal{B}}} \langle \mathcal{W}_{11} \mathcal{W}_{33} \rangle = \langle \mathcal{W}_{31} \rangle \langle \mathcal{W}_{13} \rangle,
\end{equation}
and thus the relative linear entropies coincide.

\begin{figure}
\includegraphics[scale=0.4]{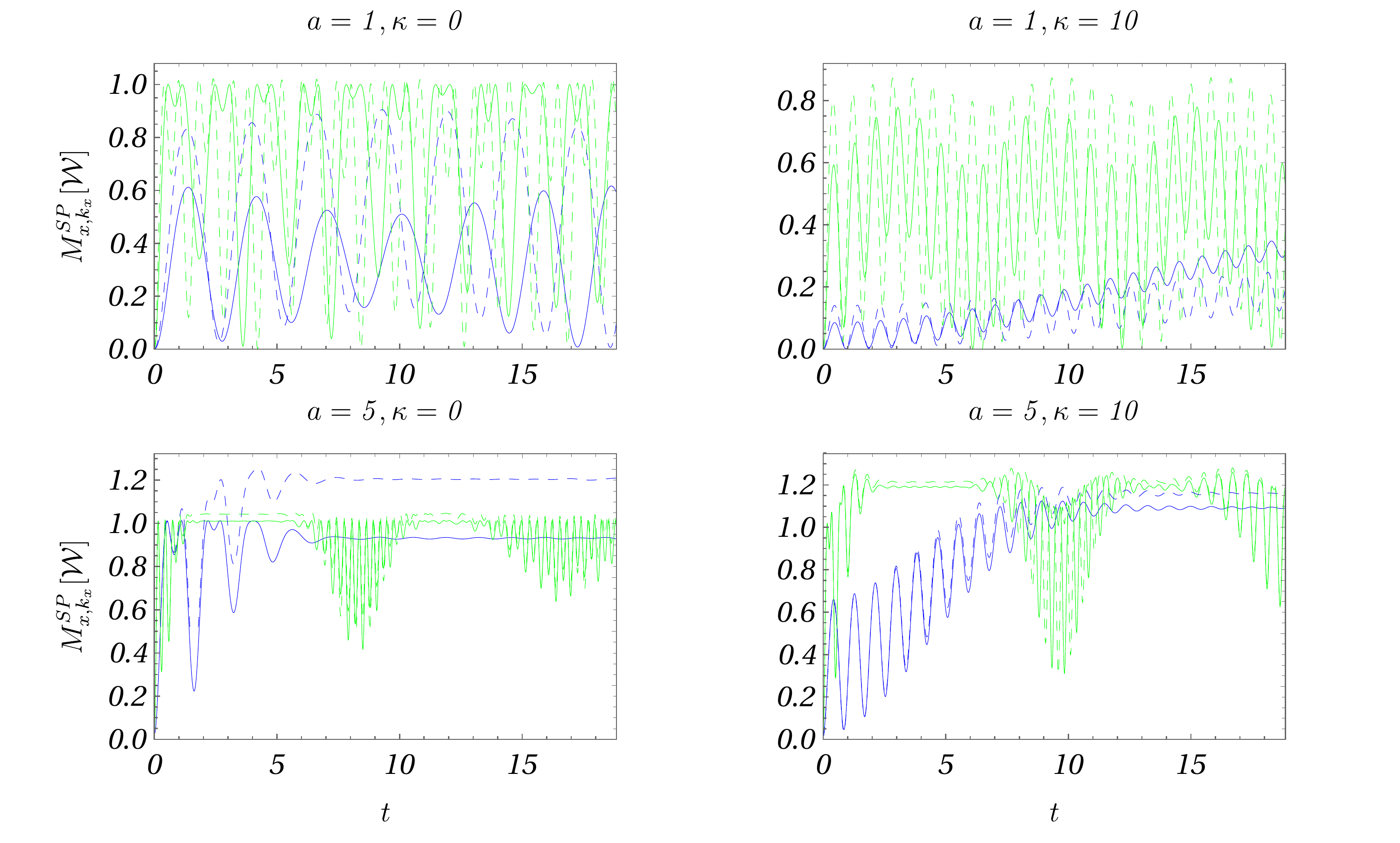}
\caption[]{Phase-space and spin-parity mutual information with parameters $m=1$, $\kappa = k_z ^2$, $\epsilon = e \mathcal{B} = 1/10$ and $1$ (blue (dark gray) and green (light gray) lines, respectively) for symmetrical (solid lines) and anti-symmetric (dashed) cat states. In the first row, the most significant contributions come from the lowest Landau levels with a definite oscillation period, which resembles the quantum information pattern found in the previous section. In the second row, correlations stagnate near their maximum value and drop off rapidly to zero when the system returns to its initial state. Such a behavior is slightly affected by increasing the magnetic field. 
}
\label{figure7}
\end{figure}

The numerical results for the mutual information (\ref{mut}) are plotted in Fig.~\eqref{figure7} for a varying distance parameter, $a$. It can be kept analytical for $a\ll1$, in which case only the first term of the series, which is $ \propto (a^2)^n$, is relevant. Otherwise, the series found in Eqs.~(\ref{eq79})-(\ref{eq81}) must be truncated.\footnote{A simple algorithm to estimate the error of truncating the series is given as follows. The series under consideration here can be generally put into the form
\begin{equation}
\sum_{n=0} ^l (...) \frac{(a^2/2)^{2n}}{(2n)!} + \sum_{n=l+1}^\infty (...)\frac{(a^2/2)^{2n}}{(2n)!} = \sum_{n=0} ^\infty (...)\frac{(a^2/2)^{2n}}{(2n)!},
\end{equation} 
where $(...)$ is smaller than unity, so let $(...) =1$ as an upper bound. While the right-hand side is simply $\cosh(a^2/2)$, the left-hand side is regarded as the series expansion of this function. The normalized error can be given as 
\begin{equation}
\mbox{Er}(a,l) =1 - \frac{ \mathcal{S}^{(l)}} {\cosh(a^2/2)},
\end{equation}
where the numerator is the finite sum truncated at $n=l$. It follows that $\mbox{Er}(a,l) = 0$ is only obtained with infinite terms; thus, a reasonable error, for instance, is $\mbox{Er}(a,l=0) \approx 0.1$ for $a = 1$ and $\mbox{Er}(a,l=8) \approx 0.1$ for $a=5$. A similar strategy applies when dealing with anti-symmetric states.} The contribution from the more excited states is to be contrasted with the results obtained for the Gaussian state preliminarily discussed. For smaller $a$, the Gaussian states interfere near the origin (cf. Fig.~\eqref{figure6}), given that only the lowest odd (even) Landau level contributes in the symmetric (anti-symmetric) case in the limit of $a \ll 1$. In this case, the averaged mutual information between spin-parity and phase space vanishes in the weak magnetic field limit, as expected from the previous results. On the other hand, for $a\rightarrow \infty$, i.e. an ideal superposition of Gaussian states, the contributions from increasing quantum numbers can be seen in the second row of Fig.~\eqref{figure7}. What stands out is that mutual information can be actually greater than unity for cat states, which confirms that spin-parity correlations indeed increase by superposing two Gaussian states. Even for a decreasing magnetic field, this behavior should be compared with the Gaussian state correlation profile, whose maximal mutual information is unity. 
\begin{figure}
	\includegraphics[scale=0.5]{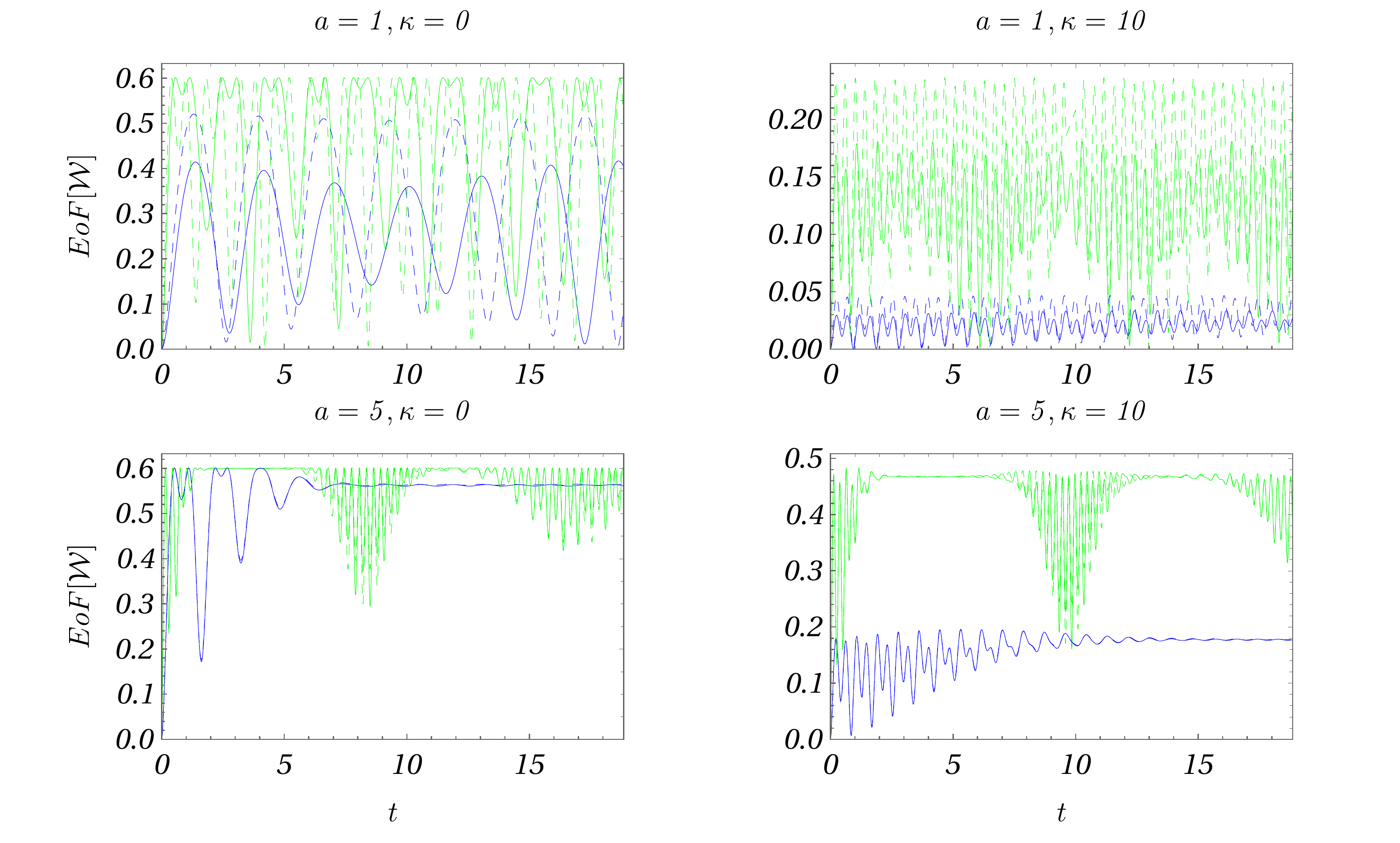}
	\caption[]{EoF with the same color scheme and parameters from Fig.~\eqref{figure7}. Quantum correlations are nearly identical between symmetrical and anti-symmetric states for increasing $a$, in which case the Dirac bi-spinor is separable only for the initial state. On the other hand, for smaller values of $a$, the oscillation pattern corresponds to the lowest Landau Levels. 
	}
	\label{figure8}
\end{figure}

There remains the question whether this behavior is also observed for quantum correlations. The phase-space averaged quantum concurrence is computed in terms of the Wigner matrix elements; it is obtained by applying Eq.~\eqref{condef} to the cat states discussed previously, which reads

\begin{eqnarray}
\langle \mathcal{C}^2 \rangle_{\{x,k_x\}} &=& - \left( \frac{2 \pi}{\sqrt{e \mathcal{B}}}\right) 2 \langle \mathcal{W}_{11} \rangle \langle \mathcal{W}_{44} \rangle \nonumber \\
&=& 8 \hspace{-1em} \sum_{ \{n,m\} \text{ odd} } \cosh(a^2/2) ^{-2} \bigg\{ \eta_n ^2 \eta_m ^2 \sin ^2(E_m t) B_m ^2 \bigg(\eta_n ^{-2} - 4(A_n ^2 + B_n ^2) \sin ^2(E_n t) \bigg) \nonumber \\
&& \hspace{20em} \times \frac{(a^2/2)^{n+m-2}}{\Gamma(n)\Gamma(m)} \bigg \},
\end{eqnarray}
over which the approximation scheme previously discussed can be applied. In particular, it holds a strong resemblance to the Gaussian state results. In fact, the $n=m$ terms correspond to the quantum concurrence for the Gaussian state computed from Eq.~\eqref{concurrence2}, once weighted by appropriate factors. 

The numerical results are shown in Fig.~\eqref{figure8}, in terms of the EoF. Due to the interference between the first Landau levels discussed above, the quantum state oscillates between non-separable and approximately separable states if $a$ is small enough. However, this oscillation is partially suppressed when there is significant contribution from states with increasing quantum number. On the one hand, EoF peaks for small values of $A_n = k_z /(E_n + m)$ at its maximum value, given that its amplitude goes with $B_n ^2 = 2n e\mathcal{B}/(E_n + m)^2 $ (cf. Eq.~\eqref{parameters}). On the other hand, a relevant aspect concerns the weak magnetic field limit, which is observed by comparing the bottom-right plots of Figs. (\ref{figure7})-(\ref{figure8}). There is an inverse trend between mutual information and EoF, which means that spin-parity classical correlations are maximized for ideal cat states if $e \mathcal{B} /k_z ^2 \ll 1$.

\section{Conclusions} \label{sec4}

Reporting about some previous results involving the $SU(2) \otimes SU(2)$ correlation profile carried by Dirac spinors \cite{extfields,diraclike01,PRB001,PRB002,PRA2018,BernardiniCP,BernardiniEPJP} Gaussian quantum superpositions for fermions trapped by magnetic fields were mapped into the Dirac-like structures and their spin-parity correlation properties driven by associated phase space variables were computed.

Considering the advantages of their mathematical manipulability, Gaussian states embedded into the Wigner-Dirac framework had their phase-space-dependent quantum-information structure examined. As noticed, due to a straightforward consequence of the Dirac equation, these states evolved into non-Gaussian phase-space configurations described in terms of Laguerre polynomials. Thus, precisely when the Wigner function could not be factorized into a product of spinorial and phase-space functions, the correlations emerged. Our results have shown that the total mutual information between phase-space and spin-parity degrees of freedom amounts to both types of $SU(2) \otimes SU(2)$ correlations, be them of classical or quantum nature, the latter being quantified by the quantum concurrence; the former being obtained via the spin-parity mutual information for the time evolved Wigner function. As obtained, the overall quantum correlations depend explicitly on the magnetic field, vanishing more quickly for weak fields. 

Lastly, symmetric and anti-symmetric superpositions of Gaussian states were also investigated.
Although the corresponding Wigner function inherits the intricacies related to the manageability of the quantum state itself (due to infinite series contributions), it still corresponds to a robust framework that can be implemented to compute quantum and classical correlations. For instance, Dirac cat Wigner functions were described by generalized Laguerre polynomials, which equivalently simplify to the Gaussian case described by Laguerre polynomials, as long as the Gaussian states are close enough in the $s$-coordinate. If the cat distance parameter, $a$, increases, the mutual information between continuous and discrete degrees of freedom can reach values greater than unity, which is unattainable for Gaussian states. In particular, even if $B_1 = \sqrt{2e\mathcal{B}}/(E_1 +m) \ll 1$, that is, for small but non-zero magnetic fields  (\ref{parameters}), classical spin-parity correlations have been noticed, whereas the EoF depicted by the quantum concurrence was strongly suppressed. In the opposite limit, however, EoF reached a maximum value for the state under consideration and classical and quantum correlations became equally relevant. This behavior is qualitatively preserved for large fluctuations of the magnetic field, with a surprising stagnation of the spin-parity correlation profile, thus indicating a stability for long periods of time as the parameter $a$ increases. Our results suggest that correlations can be manipulated not only by including external potentials on the Dirac Hamiltonian, but also by interfering stationary states through Gaussian wave packet configurations. 

To conclude, given that some previous results involving quantum information issues on Dirac-like systems  \cite{extfields,diraclike01,PRB001,PRB002,PRA2018,BernardiniCP,n001,n002,Gerritsma,Bermudez,n005,n006} have been mainly concerned with non-localized density matrices, the Wigner formalism for systems that support a Dirac-like Hamiltonian opens up a suitable scenario for studying the quantum information structure of confined fermions. For instance, it can map low energy dispersion relation platforms for both mono- and bilayer graphene which correspond to Dirac fermions that form Landau levels when undergoing a perpendicular magnetic field \cite{graph03,graph04,CastroNeto,ZBgraphene}. 
As feasible extensions, the Wigner-Dirac formalism can also be generalized so as to include thermalization effects, or even curved metric patterns, which may affect the confined Dirac spinors correlations \cite{Gallerati}, and hence deserves more investigation.

\vspace{.5 cm}
{\em Acknowledgments -- The work of AEB is supported by the Brazilian Agencies FAPESP (Grant No. 2018/03960-9) and CNPq (Grant No. 301000/2019-0). The work of CFS is supported by the Brazilian Agency CAPES (Grant No. 88887.499837/2020-00). }

\appendix 
\section{Phase-space averaging of the Wigner matrix elements}\label{appendix}
The phase-space average of squared elements of the Wigner matrix are computed by manipulating the orthonormalization relations. An application is given by
\begin{equation}
\langle \mathcal{W}_{11} ^2 (s,k_x;t) \rangle = \left( \int dx \int dk_x \, \mathcal{W}_{11} ^2 (s,k_x;t)\right),
\end{equation}
where
\begin{eqnarray} 
\mathcal{W}_{11} (s,k_x;t) = \sum_{ \{m,n \} \text{odd} } \frac{\bigg( e^{i E_{n}t} + (A^{2} _{n} + B^{2} _{n}) e^{-i E_{n}t} \bigg)}{1+ A^{2} _{n} + B^{2} _{n} } && \hspace{-0.75em} \frac{ \bigg( e^{-i E_{m}t} + (A^{2} _{m} + B^{2} _{m}) e^{i E_{m}t} \bigg)}{1+ A^{2} _{m} + B^{2} _{m} } \times \nonumber \\
&& \frac{(a/\sqrt{2})^{n+m-2}}{\sqrt{\Gamma(n)\Gamma(m)}} \mathfrak{L}_{_{(m-1) (n-1)}}(s,\,k_x).
\nonumber \end{eqnarray}
A change of notation turns out to useful, 
\begin{eqnarray}
\mathcal{W}_{11} (s,k_x;t) = \sum_{ \{m,n \} \text{odd} } C_n C^*_m \mathfrak{L}_{_{(m-1) (n-1)}}(s,\,k_x),
\end{eqnarray}
where $C_n (t) \equiv C_n$ depends on time but not on phase-space coordinates. Then, squaring the expression above and integrating,
\begin{equation}\label{eqA3}
\sum_{ \{m,n \} \text{odd} } \sum_{ \{m',n' \} \text{odd} } C_n C^*_m C_{n'} C^*_{m'} \int dx \int dk_x \, \mathfrak{L}_{_{(m-1) (n-1)}}(s,\,k_x) \mathfrak{L}_{_{(m'-1) (n'-1)}}(s,\,k_x).
\end{equation}
The orthogonality relations from Eq.~\eqref{eq76} suffice to evaluate this integral; it is simply
\begin{equation}
\int dx \int dk_x \, \mathfrak{L}_{_{(m-1) (n-1)}}(s,\,k_x) \, \mathfrak{L}_{_{(m'-1) (n'-1)}}(s,\,k_x)= \frac{\sqrt{e \mathcal{B}}}{2 \pi} \delta_{nm'} \delta_{mn'}, 
\end{equation}
which yields, by plugging it into Eq.~\eqref{eqA3},
\begin{equation}
\frac{\sqrt{e \mathcal{B}}}{2 \pi} \sum_{ \{n, m \} \text{odd} } \vert C_n \vert^2 \vert C_m \vert^2 .
\end{equation}
Therefore, the double-sum can be written as a sum squared,
\begin{equation}
\langle \mathcal{W}_{11} ^2 (s,k_x;t) \rangle = \frac{\sqrt{e \mathcal{B}}}{2 \pi}\left( \sum_{ \{n \} \text{odd} } \vert C_n \vert^2 \right)^2.
\end{equation}
Now, by integrating $\mathcal{W}_{11} (s,k_x;t)$, 
\begin{equation}
\langle \mathcal{W}_{11} (s,k_x;t) \rangle = \sum_{ \{m,n \} \text{odd} } C_n C^*_m\int dx \int dk_x \, \mathfrak{L}_{_{(m-1) (n-1)}}(s,\,k_x), 
\end{equation}
where
\begin{equation}
\int dx \int dk_x \, \mathfrak{L}_{_{(m-1) (n-1)}}(s,\,k_x) = \delta_{mn}.
\end{equation}
Then
\begin{equation}
\langle \mathcal{W}_{11} (s,k_x;t) \rangle = \sum_{ \{n \} \text{odd} } \vert C_n \vert ^2.
\end{equation}
Finally, one has
\begin{equation}
\langle \mathcal{W}_{11} ^2 (s,k_x;t) \rangle = \frac{\sqrt{e \mathcal{B}}}{2 \pi} \langle \mathcal{W}_{11} (s,k_x;t) \rangle ^2,
\end{equation}
as desired. All other identities follow from the same arguments.

\end{document}